\documentclass{aa}  

\usepackage{natbib}
\usepackage{graphicx}
\usepackage{color}
\usepackage[boxruled,linesnumbered]{algorithm2e}
\usepackage{txfonts}
\begin{document}


  \title{Novel SIMEX algorithm for autoregressive models to estimate AGN variability}


   \author{F. Elorrieta\inst{1,2,6}
         \and
         W. Palma
          \inst{1}
          \and
          S. Eyheramendy\inst{1,3,5}
          \and
          F. E. Bauer\inst{1,4,7}
          \and
          E. Camacho\inst{4}
          }

   \institute{Millennium Institute of Astrophysics, Nuncio Monse{\~{n}}or S{\'{o}}tero Sanz 100, Of 104, Providencia, Santiago, Chile\\
              \and
             Departmento de Matematicas, Facultad de Ciencia, Universidad de Santiago de Chile, Av. Libertador Bernardo O’Higgins 3663 Estacion Central, Santiago, Chile
         \and
             Faculty of Engineering and Sciences, Universidad Adolfo Ibanez, Diagonal Las Torres 2700, Peñalolen, Chile
          \and
             Instituto de Astrof{\'{\i}}sica and Centro de Astroingenier{\'{\i}}a, Facultad de F{\'{i}}sica, Pontificia Universidad Cat{\'{o}}lica de Chile, Campus San Joaquín, Av. Vicuña Mackenna 4860, Macul Santiago, Chile, 7820436
          \and
              Data Observatory (DO)
          \and
              Center for Interdisciplinary Research in Astrophysics and Space Exploration (CIRAS), Universidad de Santiago de Chile, Av. Libertador Bernardo O’Higgins 3363 Estacion Central, Santiago, Chile
          \and
              Space Science Institute, 4750 Walnut Street, Suite 205, Boulder, Colorado 80301\\
              \email{felipe.elorrieta@usach.cl}
             }


 
  \abstract
 {The origin of the variability in accretion disks of active galactic nuclei (AGN) is still an unknown, but its behavior can be characterized by modeling the time series of optical wavelength fluxes coming from the accretion disks with damped random walk (DRW) being the most popular model for this purpose. The DRW is modeled by a characteristic fluctuation amplitude $\sigma$ and damping timescale $\tau$, with the latter being potentially related to the mass and accretion rate onto the massive black hole.
 The estimation of $\tau$ is challenging, with commonly used methods such as the maximum likelihood (ML) and the least square error (LSE) resulting in biased estimators.
 This problem arises most commonly for three reasons: i) the light curve has been observed with additive noise; ii) some cadence scheme; iii) when the autocorrelation parameter is close to one. The latter is called the unit root problem, a well-known statistical issue.} 
   {In order to improve these parameter estimation procedures for estimating the parameter $\tau$, we developed a simulation-extrapolation (SIMEX) methodology in the context of time series analysis. }
   {We consider both the standard class of autoregressive processes observed at regular intervals and a recently developed class of irregularly autoregressive processes (iAR). The performance of the SIMEX estimation method was evaluated under conditions of near-unit-root behavior and additive noise through extensive Monte Carlo simulations. Furthermore, we applied this methodology to AGN light curves from the ZTF survey, assessing its accuracy in estimating the damping timescale ($\tau$).}
   {Monte Carlo experiments confirm that the SIMEX approach outperforms MLE and LSE methods, offering a reduction in estimation bias ranging from 30\% to 90\%. Real-data applications further validate the methodology, yielding better model fits and lower mean squared errors (MSE).}
   {The more accurate estimation of the damping timescale parameter can help to better understand the currently tentative links between DRW and physical parameters in AGN.}
   \maketitle
%

\section{Introduction \label{sec:intro}}

The origin of variability in accretion disks of active galactic nuclei (AGN) is still unknown, but its behavior, in the optical part of the spectrum, can be well described as a stochastic process. To model optical light curves coming from the accretion disk of AGNs, the damped random walk (DRW) process has been a popular choice, playing an instrumental role in statistically quantifying the intrinsic variability of AGN light curves \citep[][]{Kelly_2009,MacLeod_2010,2014MNRAS.439..703G,Kasliwal_2017,Burke_2021,Sanchez_Saez_2021}.
This model is characterized by two main parameters: the damping timescale $\tau$ (also called the relaxation time),  which quantifies how fast the flux variations lose memory of past values; and the variance $\sigma$, which measures the amplitude of the variation. In the power spectral density (PSD) that describes the statistical properties of AGN variability in the frequency domain,  $\tau$ is associated with the transition between white noise (with a flat slope at low frequencies) and red noise (with a slope $\propto \nu^{-2}$ at high frequencies).
 
Several authors have found correlations between $\tau$ and the physical properties of black holes such as the mass or luminosity \citep[][]{Burke_2021,Arevalo_2024_1,Arevalo_2024}. If these correlations or dependencies are confirmed across several orders of magnitude, i.e., from supermassive black holes (SMBHs) to intermediate-mass black holes (IMBHs) and from faint to bright AGNs, variability could be established as a powerful tool to find and characterize these objects. In that sense, regardless of the stochastic model used, the fit output parameters are tools that can be used to better understand the intrinsic physical mechanism behind the variability in accretion disks. Finally, it is important to point out the empirical challenge in astronomy of understanding and dealing with observational errors in which observing conditions and instrumental effects may play an important role. \\
  

MacLeod (2010), has applied the DRW model to quasar light curves, enabling the estimation of characteristic timescales and providing insights into the underlying physical processes driving quasar variability. The mentioned author’s work, has highlighted that AGN variability often exhibits red-noise behavior, where fluctuations at longer timescales dominate, and the DRW model offers a framework to quantify this behavior.\\



The accurate modeling of AGN variability remains an ongoing endeavor, with the DRW model serving as a foundational tool but not necessarily the sole solution for all cases e.g., \cite{Mushotzy_2008}. For example, Kozlowski (2010) explored AGN variability by employing structure functions, which offer another perspective on AGN light curve behavior. Another useful model for estimating the damping timescale and $\sigma$ is the {\it irregular autoregressive model} (iAR) \citep{iAR}. This model assumes discrete times as opposed to continuous times (like the DRW model) and can handle irregular gaps in observational times. The errors of the model can follow different distributions. For example, Gaussian (as the DRW), Gamma or T-Student, allowing some extra flexibility. The parameters of the iAR are denoted by $\phi$ corresponding to the autoregressive parameter, and $\sigma$, the standard deviation of the model. The parameter $\phi$ is related to the damping timescale by the formula,

\begin{equation}  \label{tau} 
\tau=-\frac{1}{\log(\phi)}.
\end{equation}

A particular challenge facing these models (DRW and iAR) consists on estimating the relaxation parameter $\tau$ as a function of the autocorrelation parameter $\phi$ of the model, when  $\phi$ is close to $1$. This happens when an observed time series remains with significant autocorrelation regardless of large observational gaps.  In statistics, this is a known problem in time series analysis that is commonly referred to as the near root estimation problem.\\

Time series parameter estimation in the near root context is a difficult problem to deal with because in this situation the process is unstable and near to nonstationary conditions. Also, the estimation procedure is conducted close to the boundaries of the parameter space [i.e. $(0,1)$] generating both theoretical and computational problems (see for example \cite{Barreto_2024} and references therein).\\

The near root problem is particularly acute in the context of irregularly observed time series, leading to biased estimators. In order to improve the parameter estimation, we propose to fit the time series with the iAR model, and correct the bias in the estimation of the parameter of the  model by an adaptation of the SIMulation-EXtrapolation (SIMEX) algorithm. The SIMEX algorithm is not new in astronomy. It has been implemented in several different contexts \citep[e.g.,][]{feigelson2012modern,Juvela_2013,Freeman_2009,Fernandez_2019,Escamilla_Rivera_2021,rana2017probing,Carstens_2017}. 

The SIMEX  algorithm was introduced and discussed by  \cite{Cook_1994}, \cite{Stefanski_1995}, and \cite{Carroll_1996} in the context of linear regression models, to correct for bias in the estimation of the parameters, when the bias was induced by measurement errors on covariates. We adapted the algorithm for the case of autoregressive models, which lack a covariate with measurement error, but where we could assume instead additive noise in the stochastic process. Additive noise arises naturally in astronomical observations with, for example, fluctuations in the precision of some instruments. We assessed the performance of the algorithm on simulated series generated from a regular autoregressive model (i.e., AR) and from an irregular autoregressive model (i.e. iAR). 

We also model the flux time series of IMBH candidates. The light curves were obtained from the Forced Photometry service of the Zwicky Transient Facility (ZTF), comprising five years of observations in the g and r-bands, with much more sparsely acquired additional i-band observations. This dataset presents two challenges, the cadence of 2-3 days, similar to the expected characteristic timescales of IMBHs, that might introduce bias in the parameter estimation, and the seasonal gaps, due to the fact that the facility is a ground-based observatory plus the observational effect of other astronomical objects like the moon, implying an irregularly observed time series.\\


The remainder of the paper is organized as follows. In Section~2  the basics of the regular and irregular autoregressive models are presented. The methodologies for parameter estimation are described in Section~3:  maximum likelihood estimation (MLE) in Section~3.1;   least squares error estimation (LSE) in Section~3.2; the general framework of estimation through the SIMEX algorithm in Section~3.3 and; the novel adaptation for autoregressive models in Section~3.4.  The finite sample performance of these estimation methods is assessed via Monte Carlo simulations, and the results are shown in Section~4. The proposed estimation methodology is presented in section 5 on real and assimilated data. We performed a data assimilation experiment, where we generated synthetic light curves of AGN objects, in order to estimate the relaxation parameter in a context where this parameter is known.  And also, we show examples of the SIMEX estimation in AGN light curves observed from the ZTF survey. In Section~6 the conclusions and final remarks of this work are presented.

\section{Models}

\subsection{Regular Autoregressive model}

A stochastic process $y_{t}$ is said to be autoregressive of order one, $AR(1)$, if it is defined by:

\begin{equation}  \label{ARModel} 
y_{1}=\sigma  \, \varepsilon_{1}, \hspace{0.3cm} y_{t}=\phi \, y_{t-1} + \sigma \, \sqrt{1-\phi^{2}}  \, \varepsilon_{t},
\end{equation}

\noindent
for $t=2,\ldots,n$, where $\varepsilon_{t}$ is a white noise sequence with zero mean and unit variance. Moreover, $y_{t}$ is said to be a weakly stationary process when $|\phi| < 1$ \citep{Brockwell_1991}. The mean and variance of the process is therefore given by

\begin{equation}
E(y_{t})=0 \mbox{ and }  Var(y_{t})=\sigma^2 \mbox{ for all } t, 
\end{equation}

\noindent
and for any two observational times $t,s$ we can define the autocovariance function as depending only on the absolute time difference,
 
\begin{equation}
\gamma(|t-s|)=E(y_t \, y_s)= \sigma^2 \,  \phi^{|t-s|},
\end{equation}

\noindent
as well as the autocorrelation function (ACF), 
$$\rho(|t-s|)=\frac{\gamma(|t-s|)}{\gamma(0)}= \phi^{|t-s|}.$$

If the $AR(1)$ process $Y_{t}$ is observed with an additive error $\xi_{t}$ with zero mean and known variance $\sigma_{\xi}^2$, the model \eqref{ARModel} can be rewritten as,

\begin{equation}  \label{ARModelwe} 
Y_{t}= y_t + \xi_{t}
\end{equation}

\noindent
where $y_t$ is the $AR(1)$ process without error defined in \eqref{ARModel} .\\

\subsection{Irregular Autoregressive model}

We denote $y_{t_j}$ as an observation measured at time $t_j$, and consider an increasing sequence of observational times $\{t_j\}$ for $j=1,\dots,n$. As defined by \cite{iAR}, the irregular autoregressive (iAR) process is:
\begin{equation}  \label{iARModel} 
y_{t_1}=\sigma  \, \varepsilon_{t_1}, \hspace{0.3cm} y_{t_j}=\phi^{t_j-t_{j-1}} \, y_{t_{j-1}} + \sigma \, \sqrt{1-\phi^{2(t_j-t_{j-1})}}  \, \varepsilon_{t_j},
\end{equation}

\noindent
 for  $j=2,\ldots,n$, where $\phi\in (0,1)$ and $\varepsilon_{t_j}$  are assumed independent Gaussian random variables with zero mean and unit variance (note that this model allows also for other distributional assumptions \citet{iAR}). Note that
\begin{equation}
E(y_{t_j})=0 \mbox{ and }  Var(y_{t_j})=\sigma^2 \mbox{ for all } y_{t_j}, 
\end{equation}

\noindent
and the covariance between $y_{t_k}$ and $y_{t_j}$ is  $E(y_{t_k} \, y_{t_j})= \sigma^2 \,  \phi^{t_k-t_j}$, for $k\geq j.$

Given the results above, the sequence $\{y_{t_j}\}$ corresponds to a second-order or weakly stationary process \citep{iAR}. Note that when the observational time gaps are equal, i.e. $\delta_j=t_j-t_{j-1}=c$, the model reduces to the standard regular autoregressive model AR \citep{Brockwell_1991}. However, astronomical light curves are nearly always observed with irregular times gap. For example, in \cite{iAR}, the authors discussed that a suitable representation for the observational times of the Vista Variable of the Via Lactea survey \citep{Minniti_2010} is given when the time gaps $\delta_j$ are defined by a mixture of two exponential distributions with rates $\lambda_1$ and $\lambda_2$, and weights $w_1$ and $w_2$, such that

\begin{equation}
f(\delta_j|\lambda_1,\lambda_2,\omega_1,\omega_2)=\omega_1 g(\delta_j|\lambda_1)+\omega_2g(\delta_j|\lambda_2) ~~~ \forall j=2,\ldots,n,
\end{equation}

\noindent
where $g(.)$ is the exponential distribution. With this mixture model, time-gaps can vary between short and longer periods without observations. Since the time gaps $\delta_j$ must be greater than zero, other distributions considered in the literature for a random generation of these time gaps are the $Exponential$, $Gamma$ or $Uniform(a,b)$ with $b>a>0$.

\section{Estimation}

To fully specify the model described above, it is required to estimate the two parameters: $\phi$ and $\sigma$. In this study we focus on the estimation of the parameter $\phi$ and assume, without loss of generality, that $\sigma=1$, which can be achieved by standardizing the sequences.

In what follows, we describe three methods for estimating the parameter $\phi$. These include the well-known MLE and LSE. The third approached considered is based on the SIMEX algorithm \citep{Stefanski_1995}. 

\subsection{Maximum Likelihood Estimation}%

A well-known method for estimating the parameter $\phi$ is {\em maximum likelihood estimation} (MLE) which consists in optimizing the likelihood function as follows:
\begin{eqnarray*}
L(\phi) &=& \log(\det (\Sigma_\phi)) + Y' \Sigma_\phi^{-1} Y\\
 \widehat{\phi}_n &=& \mbox{argmin} \; L(\phi),
\end{eqnarray*}
where, $Y = (y_{t_1},\ldots,y_{t_n})$ represents the vector of random variables that are assumed to follow an autoregressive model and $\Sigma_\phi = \left(\phi^{t_i-t_j}\right)_{i,j=i,\ldots,n}$ corresponds to its covariance matrix.

We observe that for the regular autoregressive model,  the variance of the ML estimator is explicitly  given by 

\begin{equation*}
Var(\widehat{\phi}_n)=\frac{1}{n}\left(1-\phi^2\right).
\end{equation*}

\noindent Note also that, this variance approaches $0$ as $\phi \to 1 $.

\subsection{Least Square Error Estimation}%

An alternative method for estimating the parameter $\phi$ is the {\em least square error} (LSE). In this case, the estimated parameter is obtained as the argument which minimizes the sum of squares errors defined by

\begin{eqnarray*}
J(\phi) &=& \mathop{\sum}\limits_{i=2}^{n} (y_{t_i}- \phi^{t_i-t_{i-1}} y_{t_{i-1}})^2\\
 \tilde{\phi}_n &=& \mbox{argmin} \; J(\phi),
\end{eqnarray*}

\noindent
 where $\tilde{\phi}_n$ is the least square estimator of $\phi$.

\subsection{SIMEX method}%

Before we delve into the specifics for estimating $\phi$ based on the SIMEX algorithm, we introduce in general how this algorithm works. In Section \ref{sec:simex} we present the adaptation of this algorithm for the problem in this study.

\cite{Cook_1994} introduced the SIMEX method, which has become popular among statisticians and scientists due to its simplicity and versatility (e.g., \cite{WANG201025}, \cite{li2019linear}, \cite{amico2018cure}, \cite{li2003functional} and \cite{guolo2008robust}). This method is effective for models with measurement error in the predictors when the standard errors of measurements are known or can be reasonably estimated. It yields nearly asymptotically unbiased and efficient parameter estimates for various regression models and less-biased non-regression population parameter estimates (e.g., \cite{Carroll_1996}; \cite{carroll1999nonparametric}; \cite{Cook_1994}; \cite{Kuchenhoff_2006}; \cite{piesse2008using}; \cite{Stefanski_1995}).

The SIMEX method involves adding simulated measurement error with increasing variance to the original data, applying statistical models to these increasingly error-prone datasets, identifying the trend of model parameter estimates against the variance of the added measurement error, and extrapolating this trend back to a point with no measurement error. While it doesn't provide an analytic solution to the measurement error problem and it is computationally intensive, its strength lies in its broad applicability to both simple and sophisticated measurement error models \citep{Cook_1994}. Given that standard errors of measurements are commonly estimated and reported in astronomy, the SIMEX method has potential in a wide range of applications.\\

\subsubsection{{\bf The Simulation step:}}

The SIMEX procedure begins with a simulation stage where simulated measurement errors are added to the predictor, say \( X \), making it increasingly error-prone. It is assumed that the predictor \( X \) is observed with error, i.e. $X = X_{\text{true}}+V$, where \( V \) is a random variable representing the measurement error of the predictor. In our implementation, we assume \( V \sim N(0, \sigma^2) \). These error-augmented predictors, known as re-measurements of \( X \), are created by first choosing a set of monotonically increasing small numbers, such as \( \lambda = \delta, 2\delta, \ldots, l\delta \), where $\delta$ is the constant increase in $\lambda$ in each simulation. For each \( \lambda \), an artificial error \( \sqrt{\lambda} U \) is generated, where \( U \) is a random variable, which we also assume \( N(0, \sigma^2) \). The re-measurements, \( X(\lambda) \), are then \( X(\lambda) = X + \sqrt{\lambda} U = X_{\text{true}} + V + \sqrt{\lambda} U \). The variance of the inflated measurement errors of \( X(\lambda) \) is \( (1 + \lambda) \sigma^2 \). $\sigma^2$ is considered known, or there is a way of properly estimating it.

Next, the response variable, say \( Y \), is regressed on \( X(\lambda) \), i.e. $Y=\beta_0+\beta_1X(\lambda)+\epsilon$. As \( \lambda \) increases, \( X(\lambda) \) becomes more endogenous, and the regression coefficient estimates become increasingly biased. This relationship forms the basis for the extrapolation. To reduce sampling variation, multiple re-measurements of \( X \) are generated for each \( \lambda \). The naive coefficients for each re-measured \( X \) are denoted as \( \beta_{k, \text{naive}}(\lambda, B) \), where \( k \) indicates intercept (\( k = 0 \)) or slope (\( k = 1 \)), and \( B \) is the number of re-measured predictors. The sample mean, \( \hat{\beta}_{k, \text{naive}}(\lambda) \), is calculated for extrapolation to reduce sampling error.

The choice of \( \lambda \) is crucial. \citet{Stefanski_1995} suggest restricting \( \lambda \) values to a neighborhood of zero to minimize extrapolation errors, proposing a range between 0 and 2. Different studies have used various \( \lambda \) values: \citet{carroll1999nonparametric} used \( \lambda = \frac{1}{2}, \ldots, \frac{4}{2} \) for nonparametric regression, while \citet{piesse2008using} used \( \lambda = \frac{2}{5}, \frac{4}{5}, \ldots, \frac{10}{5} \).\\ 

\subsubsection{{\bf The Extrapolation step:}}

At this point, there is a series of \( \hat{\beta}_{k, \text{naive}}(\lambda) \) corresponding to different values of \( \lambda \). The objective of the extrapolation stage is to identify the relationship between \( \beta_{k, \text{naive}}(\lambda) \) and \( \lambda \). Recall that the remeasurement of \( X \), \( X(\lambda) \), has a measurement error variance equal to \( (1 + \lambda) \sigma^2 \). Therefore, the \( \beta_{k, \text{naive}}(\lambda) \) associated with \( \lambda = -1 \) is the SIMEX estimate of the true parameter in the error-free model, denoted as \( \beta_{k, \text{simex}} \). The SIMEX estimate \( \beta_{k, \text{simex}} \) is obtained by extrapolation in the regression model learned for $\beta_{k, \text{naive}}(\lambda)$ and $\lambda$.\\

There are many possible choices for the extrapolation function. The usual candidates include the linear, the quadratic, and the rational extrapolants (Carroll et al., 2006). The linear extrapolant function is:
\[
\beta_{k, \text{naive}}(\lambda) = \gamma_{k,1} + \gamma_{k,2} \lambda + \epsilon_{k, \text{lin}} \quad (k = 0, 1),
\]
and the simple quadratic extrapolant function:
\[
\beta_{k, \text{naive}}(\lambda) = \psi_{k,1} + \psi_{k,2} \lambda + \psi_{k,3} \lambda^2 + \epsilon_{k, \text{qua}} \quad (k = 0, 1)
\]
where the \( \gamma \)s and the \( \psi \)s are the parameters of the extrapolant model to be estimated, and the \( \epsilon \)s are the model residuals. The parameters of the extrapolant models can be estimated with OLS regressions, and the conventional tools of model diagnostics can be applied to assess model fit.

After choosing the appropriate extrapolant model and estimating the model parameters, it is easy to calculate the fitted value for \( \lambda = -1 \), which is \( \hat{\beta}_{k, \text{simex}} \), and the SIMEX process is complete.\\

\subsection{Novel implementation of the SIMEX algorithm for autoregressive models of order one.\label{sec:simex}}

We consider the observations $y_{t_j}$ for $j=1,\ldots,n$, obtained at irregular time gaps $t_1,\ldots,t_n$. For the iAR model \eqref{iARModel}, if the true parameter $\phi$ is {\it close} to $1$, common statistical methods to estimate it will generate biased estimators. The reason for this problem is that the parameter lies close to the boundary of the parametric space, i.e. $(-1,1)$, which challenges the  computations. We propose to adapt the SIMEX algorithm in order to obtained better estimators.

The original observations $y_{t_1},\ldots,y_{t_n}$ are transformed by including additive noise to each of them.

\begin{equation} \label{withnoise}
Y_{t_j} =  y_{t_j} + \eta_{t_j},
\end{equation} 

\noindent
where  $\eta_{t_j}$ is a zero-mean additive noise with known variance $\sigma_{\eta}^2$. We implement the SIMEX algorithm to the iAR model on $Y_{t_j}$. Under the standard SIMEX setting, the predictor with measurement error corresponds to the previous observation $Y_{t_{j-1}}$ for $j=2,\ldots,n$. Specifically, the two steps develop as follows.

\begin{enumerate}
\item {\bf SIMULATION}: 
For each \( \lambda = \delta, 2\delta, \ldots, l\delta \) simulate:

\begin{eqnarray*} 
Y_{t_j}(\lambda) = Y_{t_j,\lambda} &=&  Y_{t_j} + \lambda \ \nu_{t_j}\\
&=&  Y_{t_j,\text{true}} + \eta_{t_j} + \lambda \ \nu_{t_j}\\
\end{eqnarray*}

\noindent
for $j=1,\ldots,n$ and $\delta\geq 0$,  where $\nu_{t_j}$ is the white noise sequence with zero mean and variance $\sigma^2_\nu$ added to the observed time series. Once $Y_{t_1,\lambda}, \ldots, Y_{t_n,\lambda}$ have been simulated, and assumed that they have been generated under the iAR model (\ref{iARModel}), an estimator $\hat{\phi}_{\lambda,i}$ for $i=1,\ldots,B$ is obtained. Then to reduce the variance, $\hat{\phi}_{\lambda}=\frac{1}{B}\sum_{i=1}^B\hat{\phi}_{\lambda,i}$ is the estimator of $\phi$ for a given $\lambda$. As $\lambda$ increases, $\hat{\phi}_{\lambda}$ becomes increasingly biased.\\

In our implementation, two scenarios are considered: (1) when the time series has additive error (e.g., in equation \eqref{withnoise}), and (2) when it does not (e.g., in equation \eqref{iARModel}). The variance of the inflated measurement errors is $((1 + \lambda) \sigma^2_\nu )$ in the presence of measurement error and $(\lambda \sigma^2_\nu)$ otherwise, since in this case $Y_{t_j} = Y_{t_j,\text{true}}$.

 
\item {\bf EXTRAPOLATION}: 
After obtaining $\hat{\phi}_{\lambda}$ for \( \lambda = \delta, 2\delta, \ldots, l\delta \), the objective is to find the relationship between $\lambda$ and $\hat{\phi}_{\lambda}$. This is done using linear regression analysis. Our implementation offers four alternatives to fit the regression. It can be a polynomial regression of degree from 1 to 4, corresponding to  "linear", "quadratic", "cubic" and "bi-quadratic" respectively. Also, our implementation allows to set different weights to the points generated from the estimates $\widehat{\phi}(\lambda)$ as follows:

\begin{itemize}
\item {\bf equal}: $w_i = 1/N ~\forall i=1,\ldots,N$  
\item {\bf exponential}: $w_i = \alpha (1 - \alpha)^{i-1}$, where $i=1,\ldots,N$ and $\alpha$ is a parameter that determines the weight of the first point. Note that for this definition, the weights decrease as the additive error variance increases.
\item {\bf invexponential}: $w_i = \alpha (1 - \alpha)^{N-i}$, where $i=1,\ldots,N$ and $\alpha$ is a parameter that determines the weight of the last point. Note that for this definition, the weights increase as the additive error variance increases.
\item {\bf gaussian}: $w_i = f_{i\delta}(l\delta/2,\sigma^2)$, where $i=1,\ldots,l$, and $f_{i\delta}(l\delta/2,\sigma^2)$ is the density of a normal distribution with mean $\mu= l\delta/2$ and variance $\sigma^2$ evaluated at the point $i\delta$. Note that for this definition we get higher weights at intermediate points.
\end{itemize}

\end{enumerate}

Once this model is estimated, the SIMEX estimator of $\phi$ is then obtained by setting $\lambda = 0$ (when there is no measurement error) or $\lambda = -1$ (when there is measurement error). The Algorithm \ref{alg:iARSimex} summarizes the simulation and extrapolation steps:\\

\begin{algorithm}
 \caption{SIMEX algorithm for autoregressive model of order one.\label{alg:iARSimex}}
     \SetAlgoLined
 Input: $B$ (number of repetitions); $N$ (number of simulations); $\delta$ (gap between points); $type$ (type of fit); $weight$ (weights for the fit).
 \BlankLine
 \For{$i=1$ \KwTo B}{
  \For{$j=1$ \KwTo N}{
    \begin{enumerate}
    \item
    Simulate $Y_{t_j}$ as the sum of the observed\\  
    time series $y_{t_j}$ with a noise $\nu_{t_j}$ generated\\ 
    from a normal distribution with variance\\ 
    $\lambda\sigma^2_\nu$ and $\lambda=\delta,2\delta,\ldots,N\delta$;\\
\item Estimate the autoregressive parameter $\hat{\phi}_{\lambda,i}$ \\for the simulated time series $Y_{t_j}$;
 \end{enumerate}}
 Obtain $\hat{\phi}{\lambda}$ by averaging over $B$ repetitions, such that:  
 $\hat{\phi}_{\lambda}=\frac{1}{B}\sum_{i=1}^B\hat{\phi}_{\lambda,i}$;}
 Fit a polynomial regression to $\hat{\phi}_{\lambda}$ using the specified $type$ and $weight$;\\
 Obtain the SIMEX estimate by setting $\lambda = 0$ or $\lambda = -1$ depending on the scenario;\\
\end{algorithm}

\begin{figure*}
\begin{minipage}{0.53\linewidth}
\includegraphics[width=0.9\textwidth]{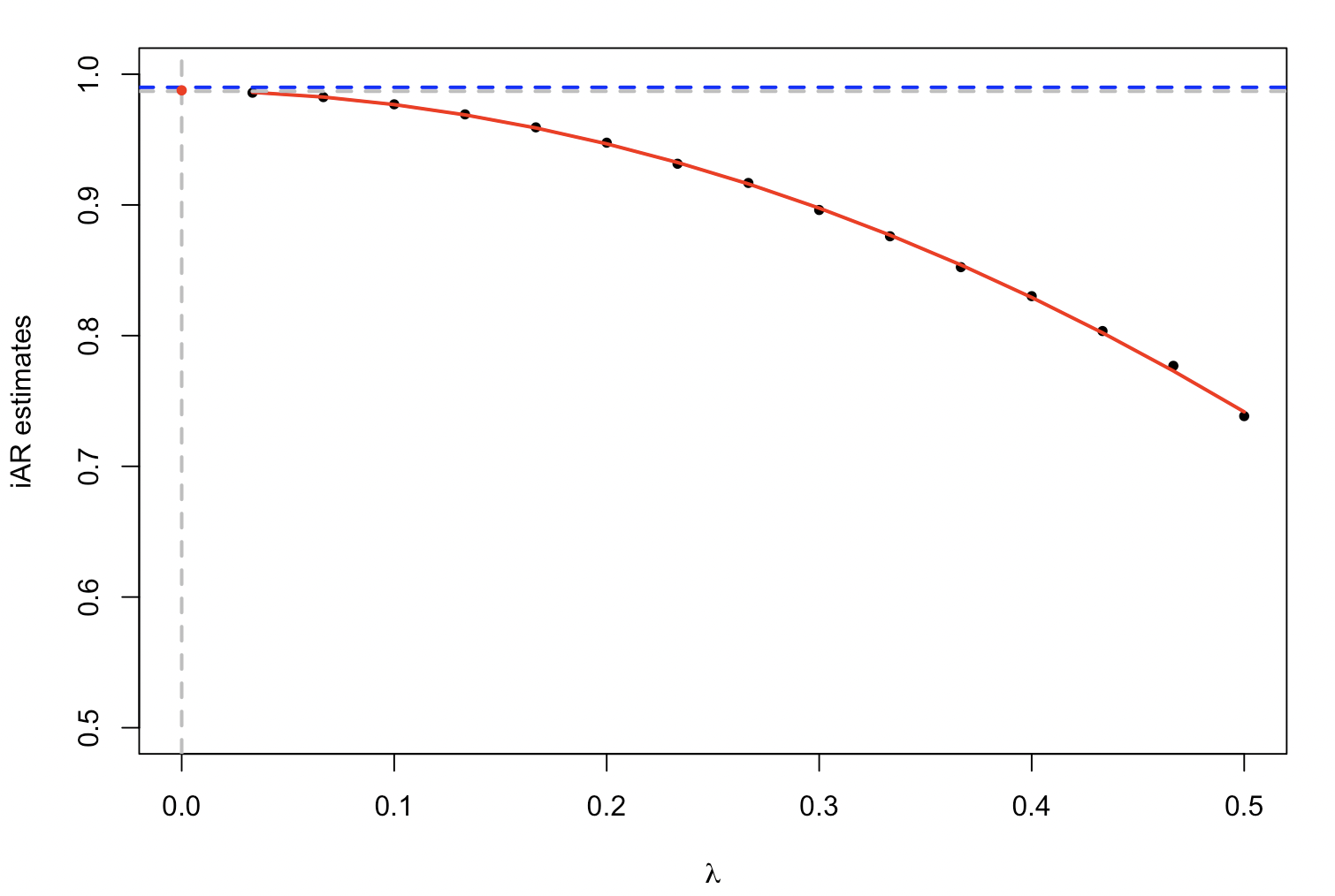}
\end{minipage}
\begin{minipage}{0.53\linewidth}
\includegraphics[width=0.9\textwidth]{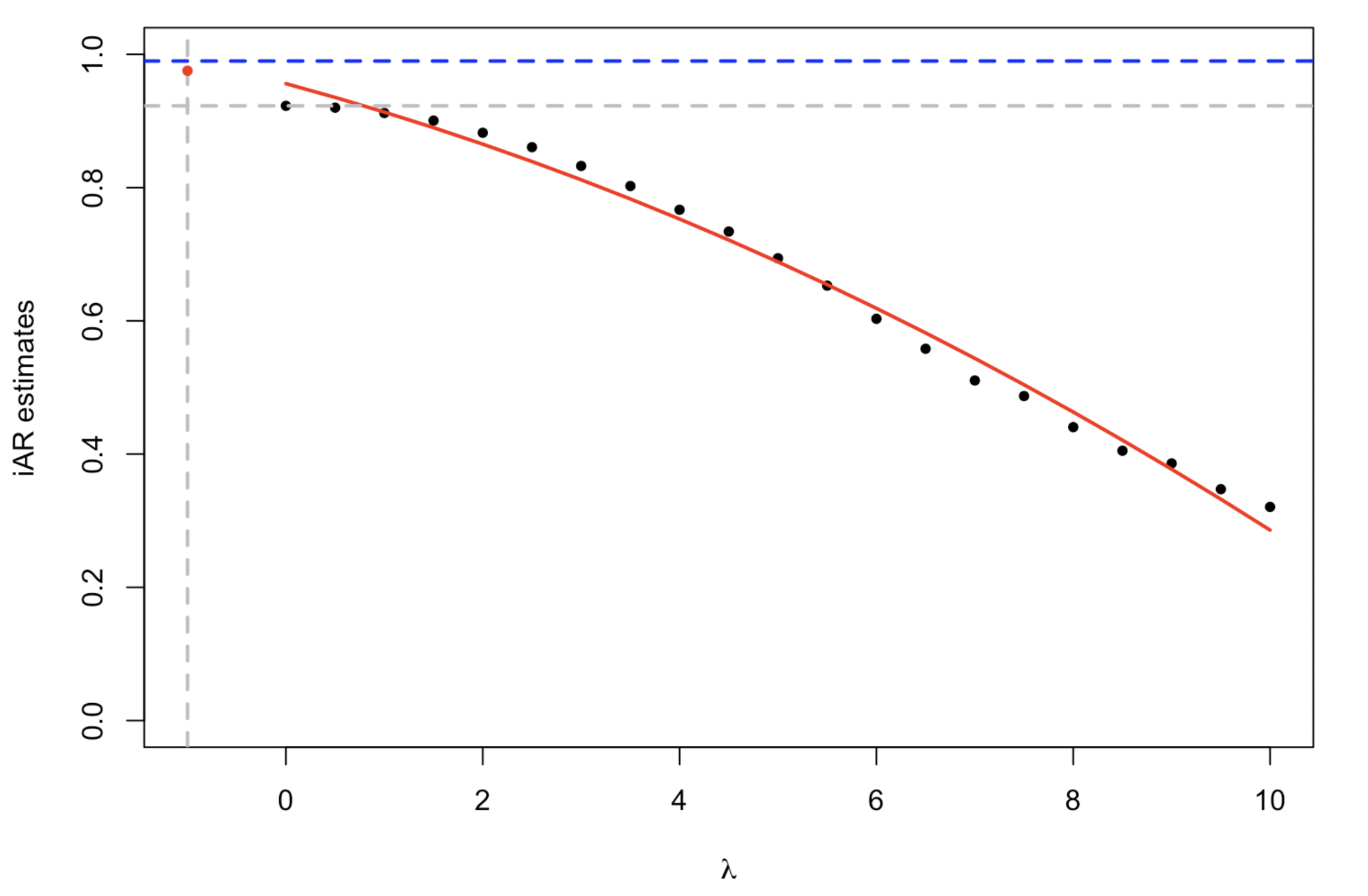}
\end{minipage}
\caption{Illustration of the SIMEX estimation algorithm. The black dots are the iAR estimates of the $\phi$ parameter when the variance measurement error is inflated by $\lambda$. The red line represents the model fitted to the black dots. The red dot indicates the SIMEX estimation of the $\phi$ parameter of the iAR model. The dashed blue line is the true value of the parameter while the horizontal dashed gray line denotes the MLE estimates. Panel (a) shows the case where the time series has no measurement error, and panel (b) illustrates the case with measurement error. For panel (a), the SIMEX algorithm parameters were: $N=15$, $B=30$, $\delta=1/30$, quadratic fit, and exponential weighting of points. For panel (b), the parameters were: $N=20$, $B=30$, $\delta=0.5$, quadratic fit, and equal weighting of points. \label{fig:illustration}}
\end{figure*}

Figure \ref{fig:illustration} illustrates the estimation process of the SIMEX algorithm for both scenarios. In panel (a), an exponential weighted quadratic fit over $N=15$ points with $\delta=1/30$ is used for a time series without measurement error. In panel (b), an equally weighted quadratic fit with $\delta=0.5$ and $N=20$ is used for a time series with measurement error variance $\sigma^2_\nu = 0.1$. Note that the main difference in the algorithm in the case that the time series has or does not have measurement error is the value of the parameter $\lambda$ at which the extrapolation is made, since in the case without error it is extrapolated to $\lambda=0$ and in the case with error it is extrapolated to $\lambda=-1$. In addition, it is important to note that in the case with small measurement error variances, larger delta values are usually required, since the aggregate error variance at each point is $\lambda\sigma^2_\nu$. Finally, in both examples presented in Figure \ref{fig:illustration} the SIMEX estimate is closer to the true parameter than the MLE estimate.\\

\section{Monte Carlo Experiments} \label{sim}

This section provides a Monte Carlo study that assesses the finite-sample performance of SIMEX, ML and LSE estimators. We fixed the parameter $\phi=0.99$ in all the experiments, and we consider two sample sizes $n=200$ and $n=500$. Each experiment was repeated $m=1000$ times. The performance criteria used to evaluate the estimation methods were the standard deviation of the estimated parameters and the difference between the estimated and true parameter, measured by the bias and the root mean square error (RMSE). Tables \ref{t1}-\ref{t2} show results for the regular AR model, and Tables \ref{t3}-\ref{t4} shows the results for the iAR model. Table \ref{t1} shows results for the standard AR model. Table \ref{t2} shows results for the AR model with additive noise with standard deviation equal to $0.1$. In addition, Tables \ref{t3}-\ref{t4} show results for the iAR model. Observational times were generated from two distributions, one that guarantees small time gaps such as Uniform$(0,1)$ and another with larger gaps such as Gamma$(3,1)$. The parameter settings for the SIMEX algorithm for all the experiments considered were $N=15$, $B=30$, linear fit and points weighted equally. The only parameter that changes between experiments is $\delta$ which is defined as $\delta=1/50$ for the regular spacing case; $\delta=1/70$ for the gamma gaps case; $\delta=1/100$ for the uniform time gap case; and $\delta=0.3$ for the additive noise case.\\

The Monte Carlo experiments show that the SIMEX estimator generally outperforms the LSE and ML estimators in estimating the autoregressive parameter $\phi$. In scenarios with no additive noise and regular observational times (Table \ref{t1}), SIMEX has the lowest bias and RMSE, especially as the sample size increases from $n=200$ to $n=500$, indicating high accuracy and precision. When additive noise ($\sigma_{\nu}=0.3$) is introduced (Table \ref{t2}), SIMEX maintains superior performance, with lower bias and RMSE than LSE and ML across both sample sizes. Under irregular observational times (Tables \ref{t3}-\ref{t4}), SIMEX continues to demonstrate robustness, particularly with observation times drawn from a Gamma$(3,1)$ distribution, where it shows minimal bias and low RMSE. Even under the more complex scenario with Uniform times, SIMEX consistently provides the most accurate and stable estimates, highlighting its reliability in challenging data conditions.  Note that, across scenarios, SIMEX reduces bias by approximately 30\%-90\% compared to LSE and ML, with the maximum reduction occurring at uniform times for a sample size of $n = 500$, while RMSE is reduced by up to 60\% in the experiment with additive noise and sample size $n = 500$.\\

\begin{table*}[ht]
\centering
\caption{AR estimation of the $\phi$ parameter.\label{t1}}
\begin{tabular}{rrrrrrrr}
  \hline
& n & Parameter & Estimate & Bias & SD & Theoretical SD (*) & RMSE \\ 
  \hline
SIMEX & 200 & 0.990 & 0.982 & -0.008 & 0.021 & 0.010 & 0.022 \\ 
  LSE & 200 & 0.990 & 0.965 & -0.025 & 0.024 & 0.010 & 0.035 \\ 
  ML & 200 & 0.990 & 0.965 & -0.025 & 0.022 & 0.010 & 0.033 \\ 
  \hline
  \hline
SIMEX & 500 & 0.990 & 0.996 & 0.006 & 0.008 & 0.006 & 0.010 \\ 
  LSE & 500 & 0.990 & 0.981 & -0.009 & 0.011 & 0.006 & 0.014 \\  
  ML & 500 & 0.990 & 0.981 & -0.009 & 0.010 & 0.006 & 0.014 \\ 
   \hline
\end{tabular}
\end{table*}

\begin{table*}[ht]
\centering
\caption{AR estimation of the $\phi$ parameter when the autoregressive model is generated with an additive noise with a standard deviation $\sigma_{\nu}=0.3$. (*) Theoretical SD is under regular times and without additive noise.\label{t2}}
\begin{tabular}{rrrrrrrr}
  \hline
 & n & Parameter & Estimate & Bias & SD & Theoretical SD (*) & RMSE \\ 
  \hline
SIMEX &    200 & 0.990 & 0.977 & -0.013 & 0.033 & 0.010 & 0.036 \\ 
  LSE &    200 & 0.990  & 0.940 & -0.050 & 0.039 & 0.010 & 0.063 \\ 
  ML &    200 & 0.990 & 0.940 & -0.050 & 0.038 & 0.010 & 0.062 \\ 
   \hline
  \hline
SIMEX &    500 & 0.990 & 0.993 & 0.003 & 0.014 & 0.006 & 0.014 \\ 
  LSE &    500 & 0.990 & 0.960 & -0.030 & 0.021 & 0.006 & 0.036 \\ 
  ML &    500 & 0.990 & 0.960 & -0.030 & 0.021 & 0.006 & 0.036 \\ 
   \hline
\end{tabular}
\end{table*}

\begin{table*}[ht]
\centering
\caption{iAR estimation of the $\phi$ parameter with observational times from Gamma$(3,1)$ distribution. (*) Theoretical SD is under regular times and without additive noise. \label{t3}}
\begin{tabular}{rrrrrrrr}
  \hline
 & n & Parameter & Estimate & Bias & SD & Theoretical SD (*) & RMSE \\ 
  \hline
SIMEX & 200 & 0.990 & 0.987 & -0.003 & 0.008 & 0.010 & 0.009 \\ 
  LSE & 200 & 0.990 & 0.982 & -0.008 & 0.009 & 0.010 & 0.012 \\
  ML & 200 & 0.990 & 0.982 & -0.008 & 0.009 & 0.010 & 0.011 \\ 
   \hline
  \hline
SIMEX & 500 & 0.990 & 0.992 & 0.002 & 0.004 & 0.006 & 0.005 \\ 
  LSE & 500 & 0.990 & 0.987 & -0.003 & 0.005 & 0.006 & 0.006 \\ 
  ML & 500 & 0.990 & 0.987 & -0.003 & 0.004 & 0.006 & 0.005 \\ 
   \hline
\end{tabular}
\end{table*}

\begin{table*}[ht]
\centering
\caption{iAR estimation of the $\phi$ parameter with observational times from Uniform$(0,1)$ distribution. (*) Theoretical SD is under regular times and without additive noise. \label{t4}}
\begin{tabular}{rrrrrrrr}
  \hline
 & n & Parameter & Estimate & Bias & SD & Theoretical SD (*) & RMSE \\ 
  \hline
SIMEX & 200 & 0.990 & 0.967 & -0.023 & 0.041 & 0.010 & 0.047 \\ 
  LSE & 200 & 0.990 & 0.942 & -0.048 & 0.043 & 0.010 & 0.065 \\ 
  ML & 200  & 0.990 & 0.942 & -0.048 & 0.039 & 0.010 & 0.062 \\ 
   \hline
  \hline
SIMEX & 500 & 0.990 & 0.991 & 0.001 & 0.028 & 0.006 & 0.028 \\   
  LSE & 500 & 0.990 & 0.971 & -0.019 & 0.020 & 0.006 & 0.028 \\  
  ML & 500 & 0.990 & 0.971 & -0.019 & 0.019 & 0.006 & 0.027 \\ 
   \hline
\end{tabular}
\end{table*}

\begin{figure}
\begin{center}
\includegraphics[scale=0.4]{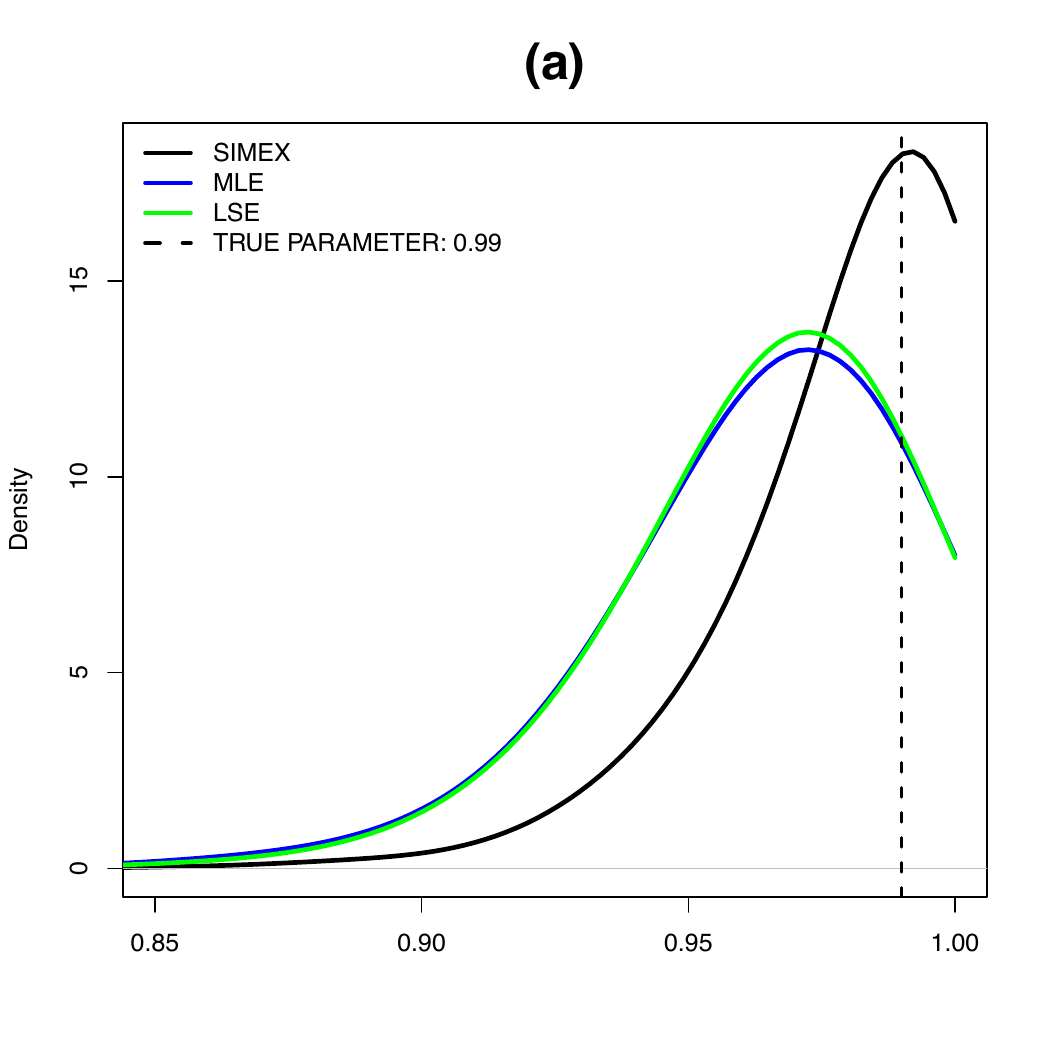}
\includegraphics[scale=0.4]{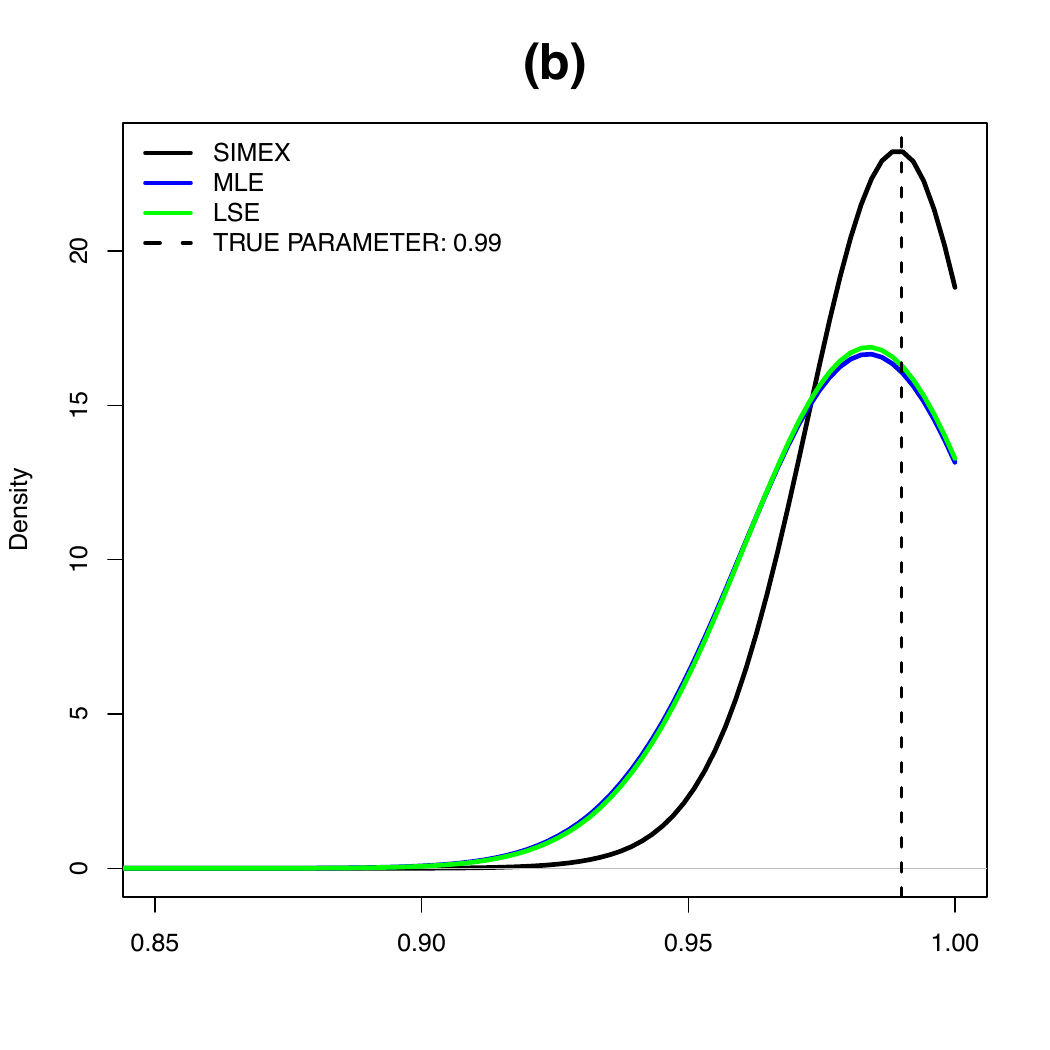}
\includegraphics[scale=0.4]{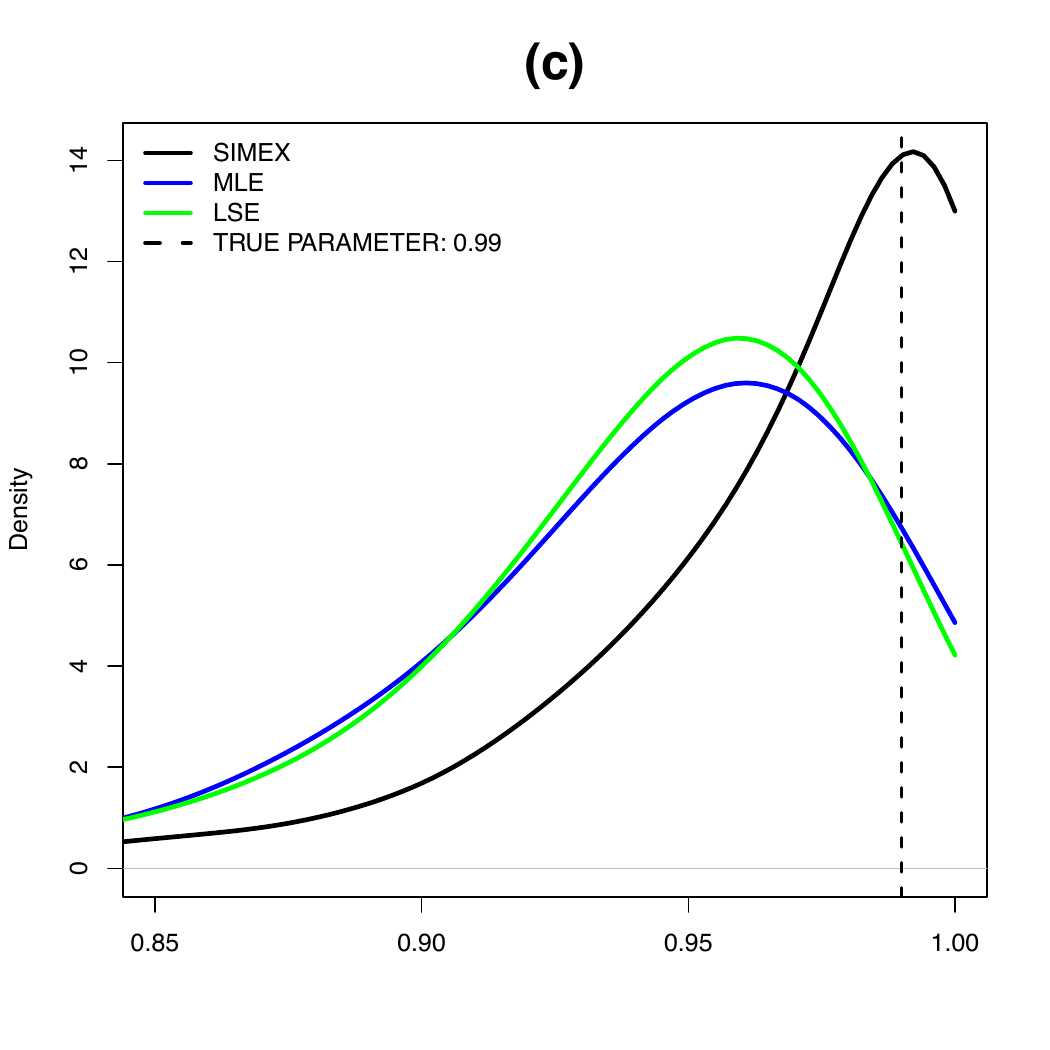}
\caption{Density of the SIMEX, ML and LSE estimators for the autoregressive paramater $\phi$ for a sample size $n=200$. Figure (a) shows the density of the estimators for the regular AR model of order 1. Figure (b) shows the density of the estimators for the iAR model simulated with a Gamma(3,1) distribution for the observational times. Figure (c) shows the density of the estimators for the iAR model simulated with a Uniform(0,1) distribution for the observational times. The dashed vertical line represents the true value of the parameter $(\phi=0.99)$.\label{fig:SimexMC}} 
\end{center}
\end{figure}

To enhance the discussion of our Monte Carlo study results, we included in Figure \ref{fig:SimexMC} three density plots that illustrate the distributional behavior of the three estimation methods (SIMEX, LSE, and ML) for simulated iAR process of size $n=200$ considering $regular$, $Gamma$ and $uniform$ time gaps respectively. These plots visually emphasize the comparative performance by showing the concentration and spread of the parameter estimates across simulations. For SIMEX, the density is notably sharper and more centered around the true parameter value, highlighting its reduced bias and higher accuracy. These density plots corroborate the findings from our simulation experiments, further demonstrating SIMEX's robustness and reliability in finite-sample settings.\\

\section{Estimation of the damping timescale in AGNs from ZTF}


In the introduction we established the relationship between the damping timescale $\tau$, or relaxation parameter and the autocorrelation parameter $\phi$ (equation \ref{tau}). Note that this equation implies the following,

\begin{equation} 
\frac{d\tau}{\tau} = \tau \, \frac{d\phi}{\phi}.
\end{equation}

This last equation shows that a small estimation error in $\phi$, say 1\%, will translate to a huge estimation error in the time relaxation parameter. For example, if $\tau=100$, then the estimation error of this parameter becomes $100\%$.\\

\begin{figure}[h!]
\centering
\includegraphics[scale=0.35]{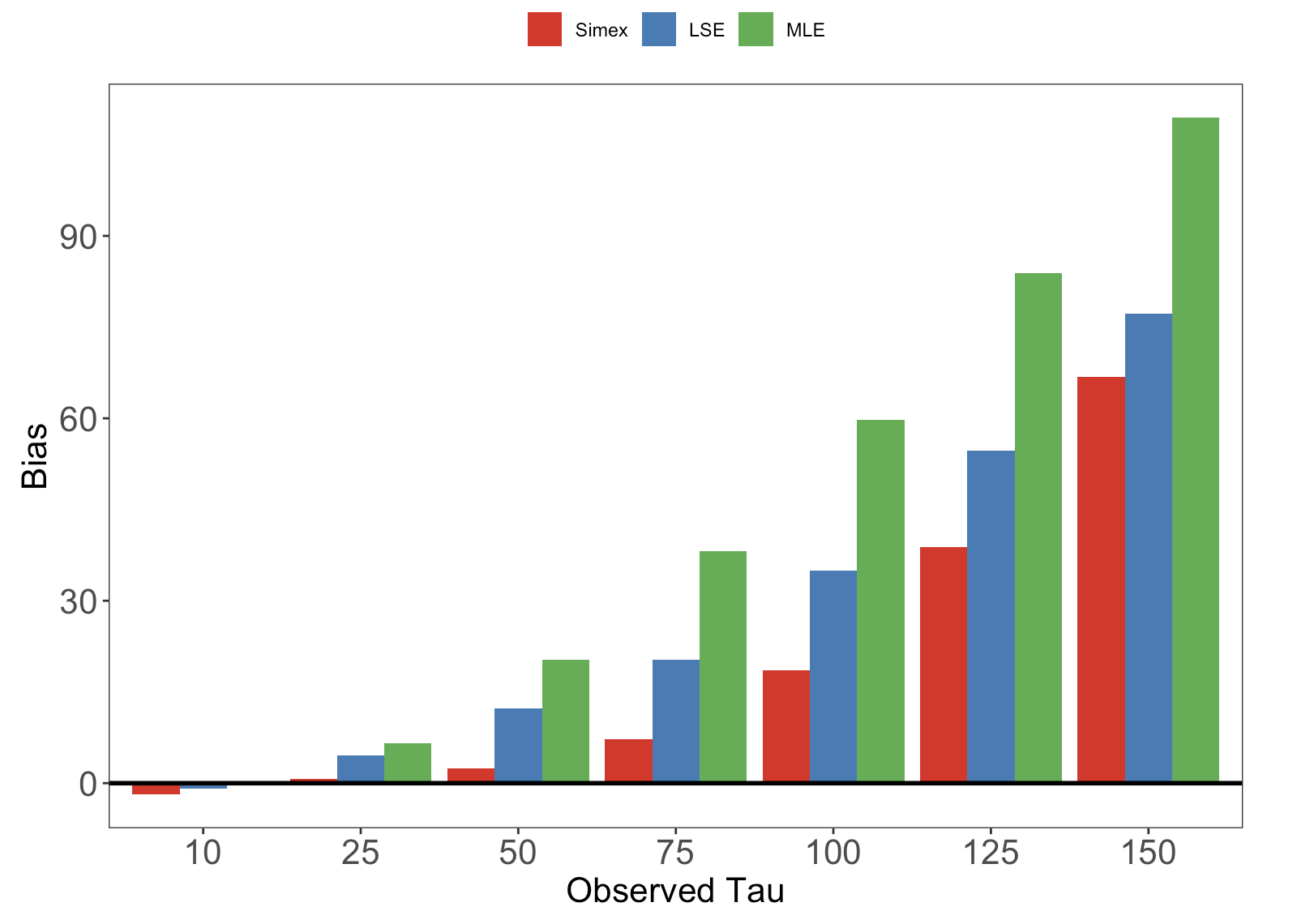}
\caption{Mean of the bias in the estimation of the parameter $\tau$ performed by the MLE, LSE and SIMEX methods for the iAR model in simulated AGN light curves. The blue, green and red bar denote the mean estimations obtained by the MLE, LSE and SIMEX methods respectively. \label{fig:Ex1AGN}}
\end{figure}

In a first experiment, we assess the estimation of the parameter $\tau$ by the iAR model using the MLE, LSE and SIMEX methods in simulated AGN light curves generated by assuming a damped random walk (DRW) model with a given $\tau$. Synthetic light curves were simulated following a similar approach as in \cite{Sanchez_Saez_2018}, i.e. adopting the exact observational times from an example long-duration, high cadence optical light curve of this reference. Specifically, this time sampling considers $175$ observed epochs over a period of 1659 days. Using this sampling, we generated $N=100$ synthetic AGN light curves for values of $\tau$ from 10 to 100. 
As in the Monte Carlo experiments, we use as the parameter set of the SIMEX estimation method, linear and equally weighted fits, $N=30$ and $\delta=1/70$. Figure \ref{fig:Ex1AGN} shows the mean bias in the estimations of the autoregressive parameter for the simulated AGN light curves across different values of $\tau$. Note in this experiment that, although the bias increases for all methods as $\tau$ becomes larger, the bias is consistently lower for the SIMEX method compared to the conventional methods (MLE and LSE). Notably, for the SIMEX method, the bias remains below 10\% relative to the true value up to a $\tau$ of 75. Specifically, at $\tau = 150$, the SIMEX method demonstrates a 40\% reduction in bias compared to the MLE method.\\

\subsection{ZTF Light Curves Example}

To assess the performance of the proposed estimation methodology on real data, we analyzed a random sample of light curves from stochastic objects, including active galactic nuclei (AGN), quasi-stellar objects (QSO), and Blazars, observed by the ZTF survey. These light curve data were processed by the ALeRCE broker \citep{Forster_2021}. The dataset consisted of 3116 light curves, each with at least 30 measurements observed in the g-band. Among these, 1609 correspond to AGN objects, 220 to Blazars, and 1049 to QSOs.\\ 

For these light curves, the damping timescale parameter $\tau$ is unknown. To determine which estimation method provides the most accurate estimate of $\tau$, we evaluate the fit of the iAR model derived from each method estimate. As goodness-of-fit measure of the iAR model we use the mean squared error (MSE) defined as,\\

\begin{eqnarray*}
\text{MSE} &=& \frac{1}{n}\mathop{\sum}\limits_{j=1}^{n}\left(y_{t_j}-\hat{y_{t_j}}\right)^{2}.\\
\end{eqnarray*}
 
For each light curve, the iAR model parameters were estimated using three methods: ML, LSE and the SIMEX estimation method proposed in this work. For the SIMEX method, four parameter settings were used, reflecting the scenarios with (Scenario 1) and without measurement error (Scenario 2). In all settings, the number of repetitions for the estimation procedure was fixed at $B=30$. To summarize:

\begin{itemize}
    \item \textbf{Setting 1:} Under Scenario 1 (no measurement error), and assuming an equally weighted quadratic fit over $N=15$ points with $\delta=1/80$.
    \item \textbf{Setting 2:} Under Scenario 1, and assuming an equally weighted quadratic fit over $N=15$ points with $\delta=1/110$.
    \item \textbf{Setting 3:} Under Scenario 2 (with measurement error), and assuming an equally weighted linear fit over $N=15$ points with $\delta=0.05$ and measurement error variance $\sigma_{\nu}=0.1$.
    \item \textbf{Setting 4:} Under Scenario 2, and assuming an equally weighted quadratic fit over $N=15$ points with $\delta=0.1$ and measurement error variance $\sigma_{\nu}=0.1$.
\end{itemize}

Among these four parameter settings, the one that minimized the MSE was selected for each light curve. Based on this goodness-of-fit criterion, Setting 3 was the most frequently chosen, accounting for 35.5\% (1,106) of the examples. Among the remaining light curves, Setting 1 was selected in 24.9\% (777) of the cases, Setting 2 in 19.4\% (603) of the cases, and Setting 4 in 20.2\% (630) of the cases.\\

Based on the SIMEX estimation setting selected for each light curve, 343 examples were identified with a near-root situation, characterized by $\phi$ estimated values greater than $0.99$, or equivalently, $\tau$ estimates exceeding $99.5$. In contrast, using the standard maximum likelihood estimation (MLE) method, 264 light curves were identified with near-root estimations.\\

For light curves where the estimation of $\phi$ is not close to the unit root, it is expected that the estimation methods should yield similar results. However, this holds true only for the SIMEX and MLE methods. As shown in Table \ref{t6}, the correlation between these two methods for non-near-root light curves is exceptionally high at $0.988$. In contrast, the LSE method shows weak correlations with both SIMEX and MLE. Considering this result and the previously observed limitations of the LSE method in simulation experiments,  we focus exclusively on MLE and SIMEX to identify light curves with near-root estimations.\\

\begin{table}[ht]
\centering
\caption{Pearson correlation between the estimations of the $\phi$ parameter obtained with the SIMEX, MLE and LSE methods in light curves that are not near-root examples.\label{t6}}
\begin{tabular}{lr}
  \hline
Methods & Correlations \\ 
  \hline
SIMEX and MLE & 0.988 \\ 
SIMEX and LSE & 0.257 \\ 
MLE and LSE & 0.296 \\ 
   \hline
\end{tabular}
\end{table}

\begin{figure*}
\begin{minipage}{0.49\linewidth}
\includegraphics[width=\textwidth]{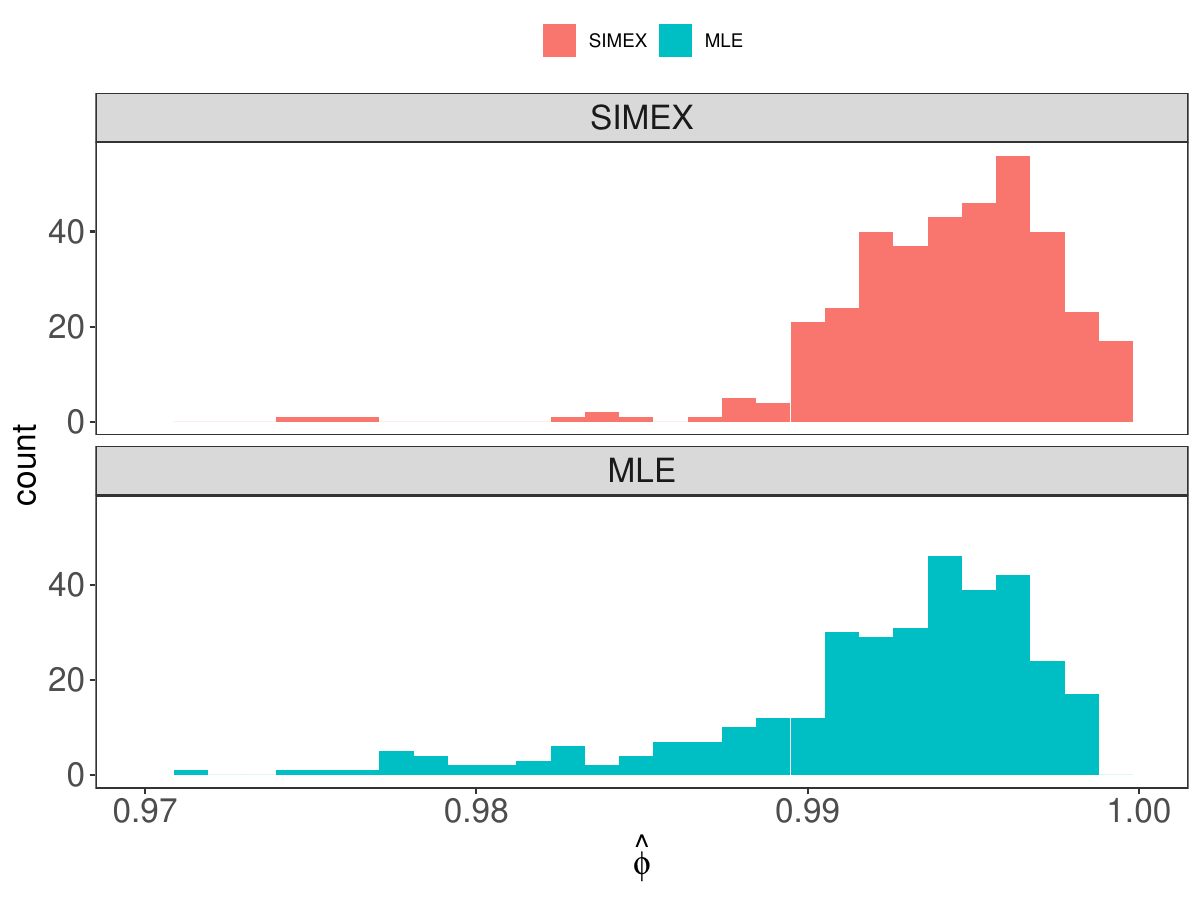}
\end{minipage}
\begin{minipage}{0.49\linewidth}
\includegraphics[width=\textwidth]{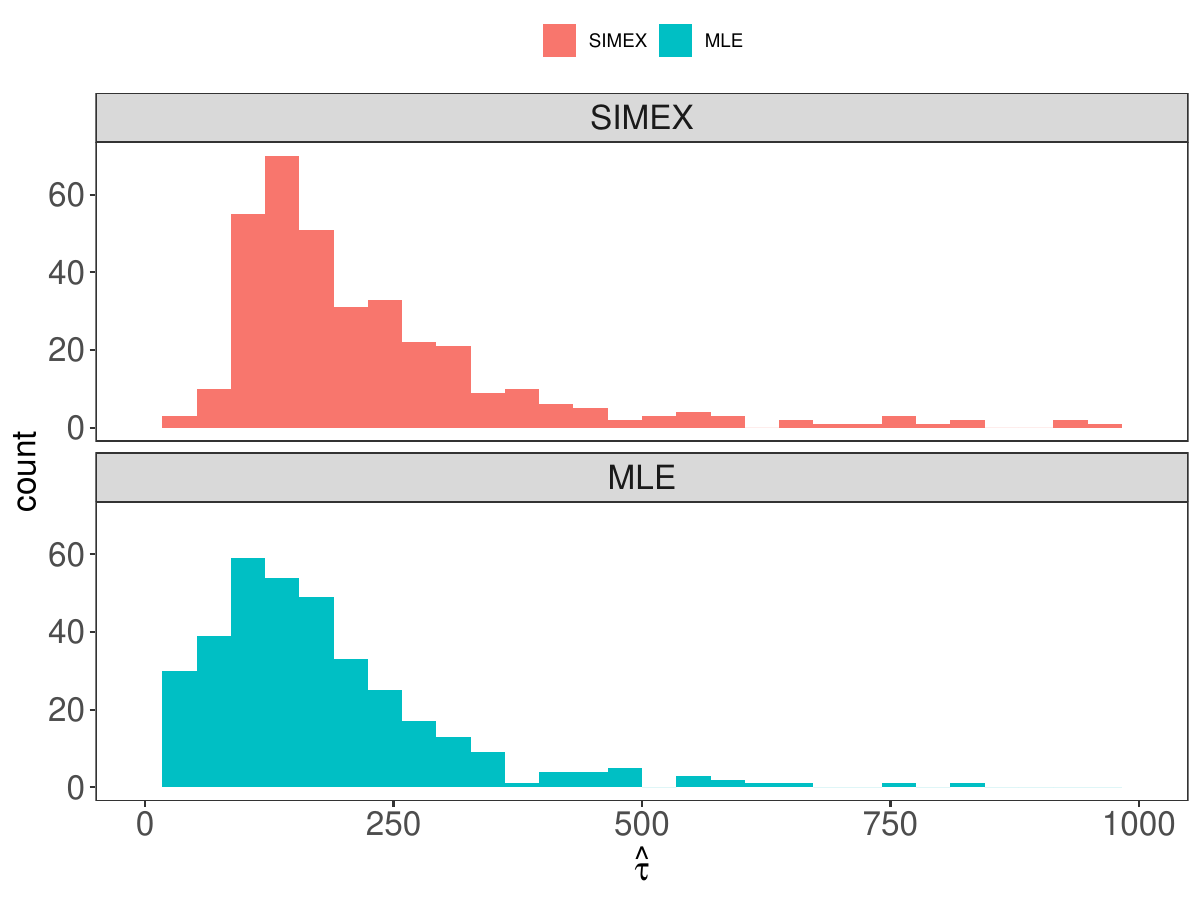}
\end{minipage}
\caption{Distribution of the estimated values of the parameters $\phi$ and $\tau$ using both SIMEX and MLE estimation methods in 365 near root light curves observed in the g-band of the ZTF survey.  \label{fig:ZTFdistribution}} 
\end{figure*}

There are 365 light curves with estimates close to the unit root in at least one of the two methods under consideration, either MLE or SIMEX. Focusing on these cases, we observe that the SIMEX method achieves the best fit, as determined by the criterion of the lowest MSE, in 318 of these examples (87.1\%). In contrast, the MLE method produces a lower MSE in 47 cases (12.9\%). Furthermore, looking at the parameters estimators for $\phi$ and $\tau$, we notice that the distribution of these estimates tends to concentrate at higher values when the SIMEX method is applied, as can be seen in Figure \ref{fig:ZTFdistribution}. For example, in terms of $\tau$, the third quartile of the SIMEX estimates is $288.98$, substantially higher than the $220.01$ for MLE, reinforcing the tendency of SIMEX to produce larger $\tau$ estimates compared to the standard maximum likelihood method.\\ 

\begin{figure*}
\begin{minipage}{0.246\linewidth}
\includegraphics[width=\textwidth]{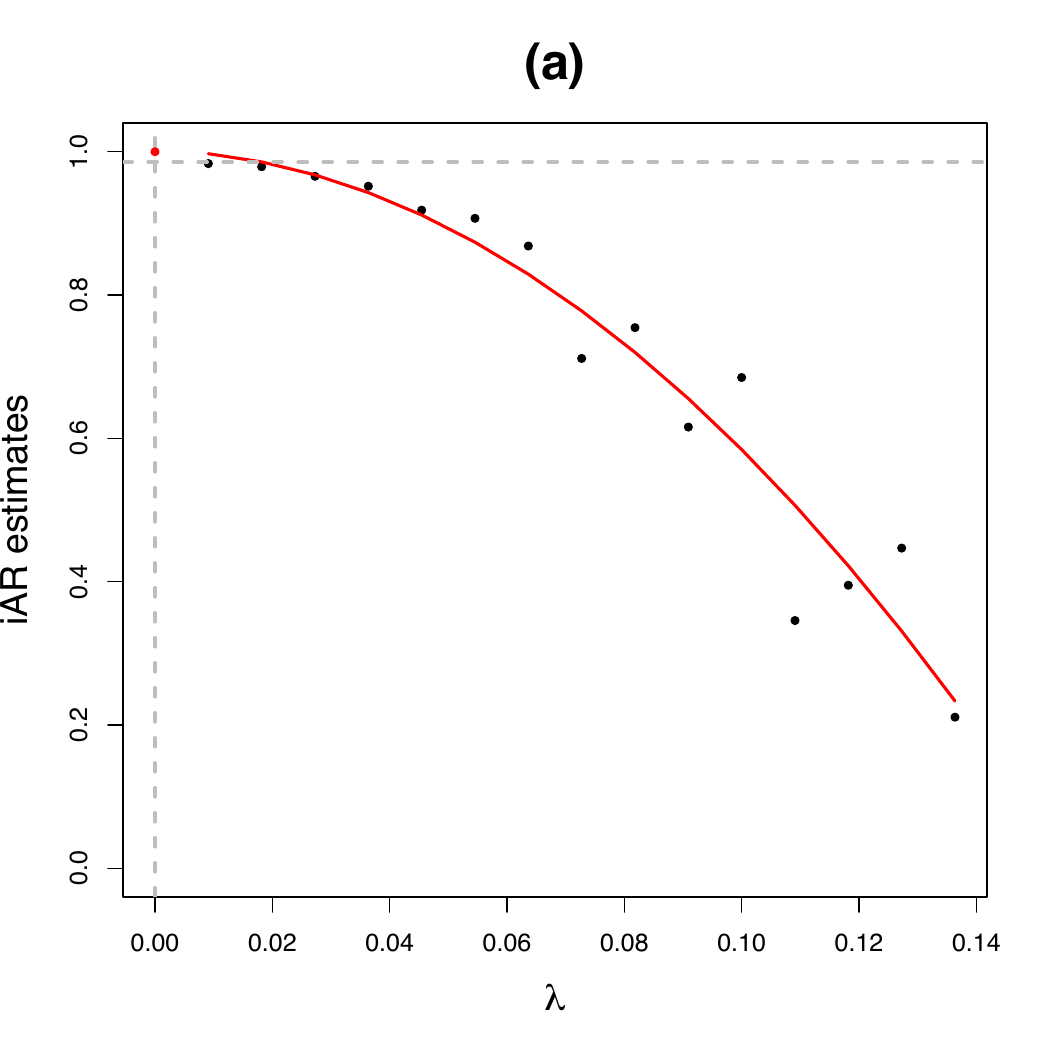}
\end{minipage}
\begin{minipage}{0.246\linewidth}
\includegraphics[width=\textwidth]{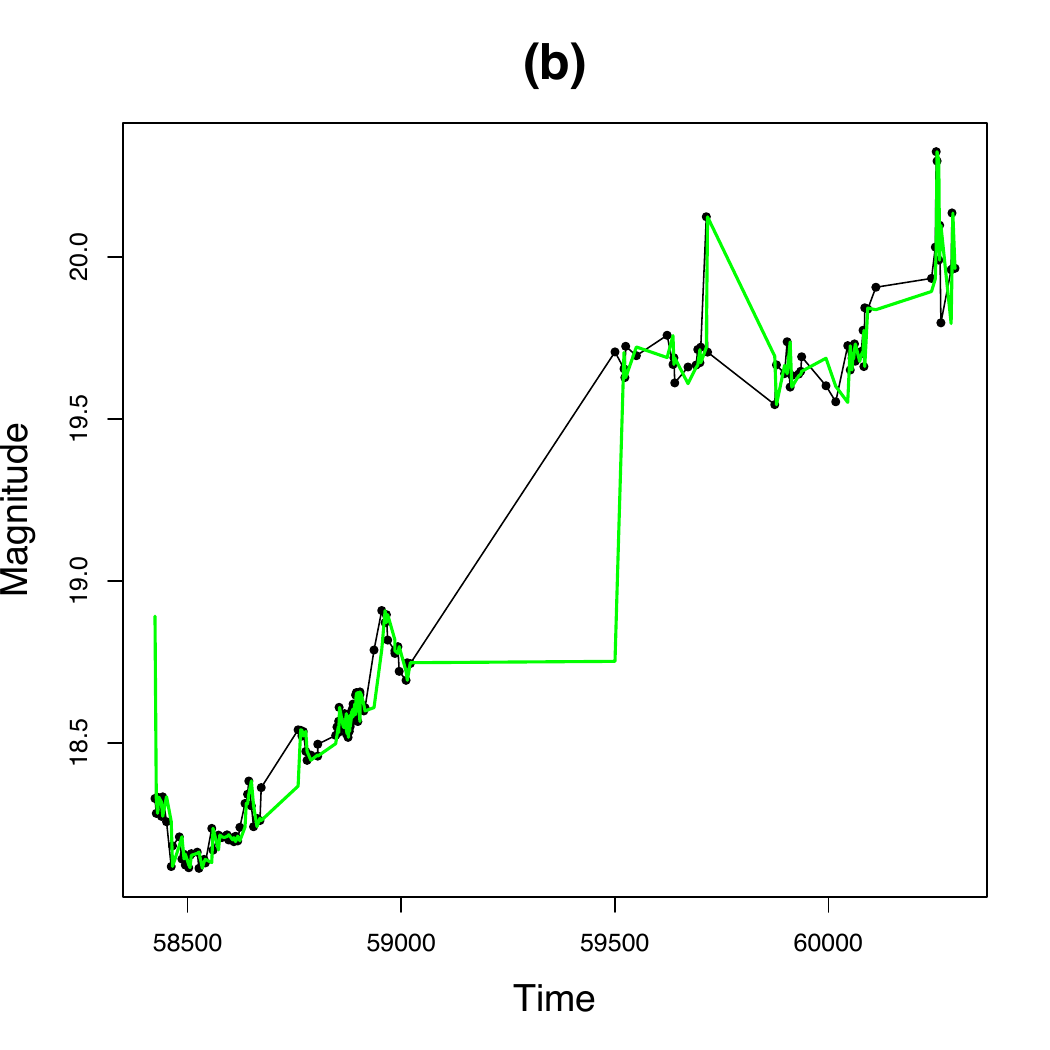}
\end{minipage}
\begin{minipage}{0.246\linewidth}
\includegraphics[width=\textwidth]{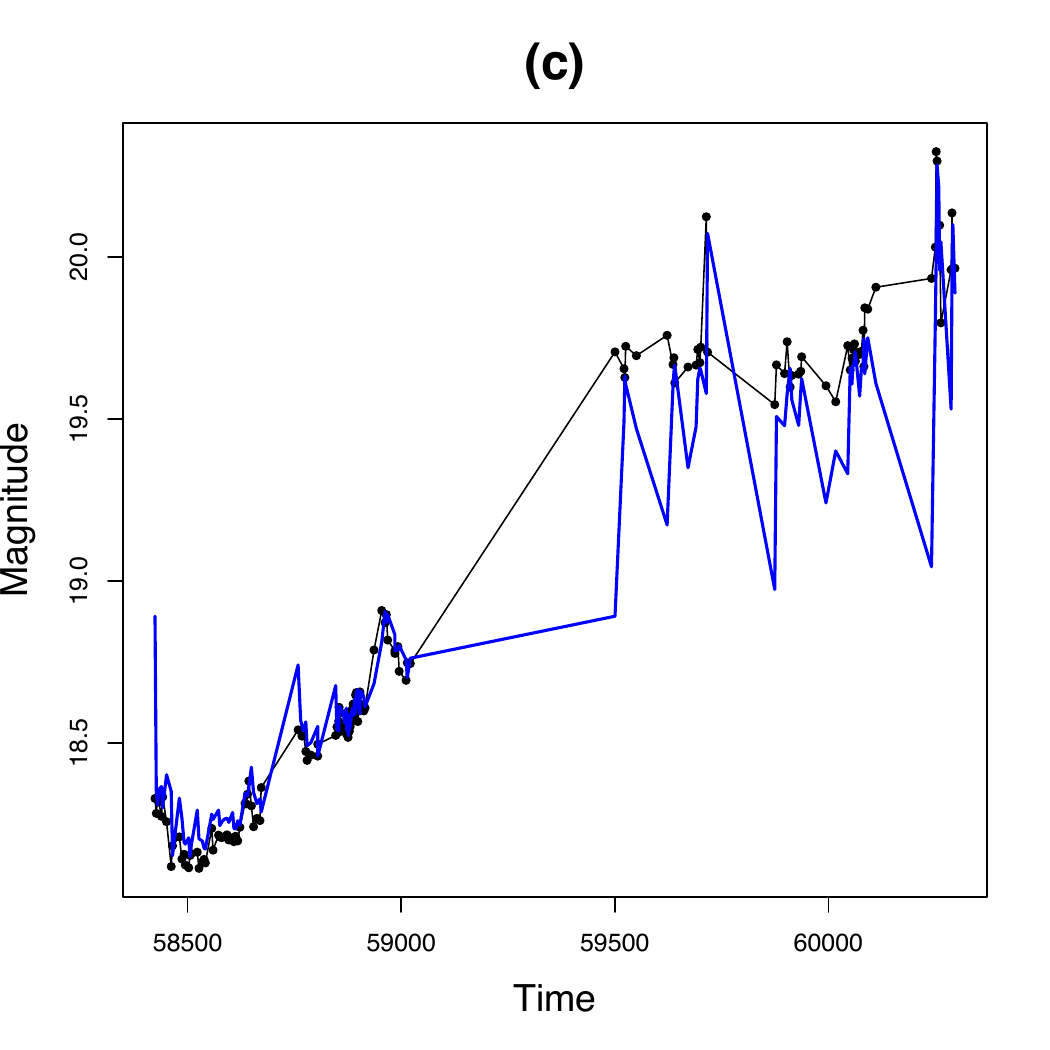}
\end{minipage}
\begin{minipage}{0.246\linewidth}
\includegraphics[width=\textwidth]{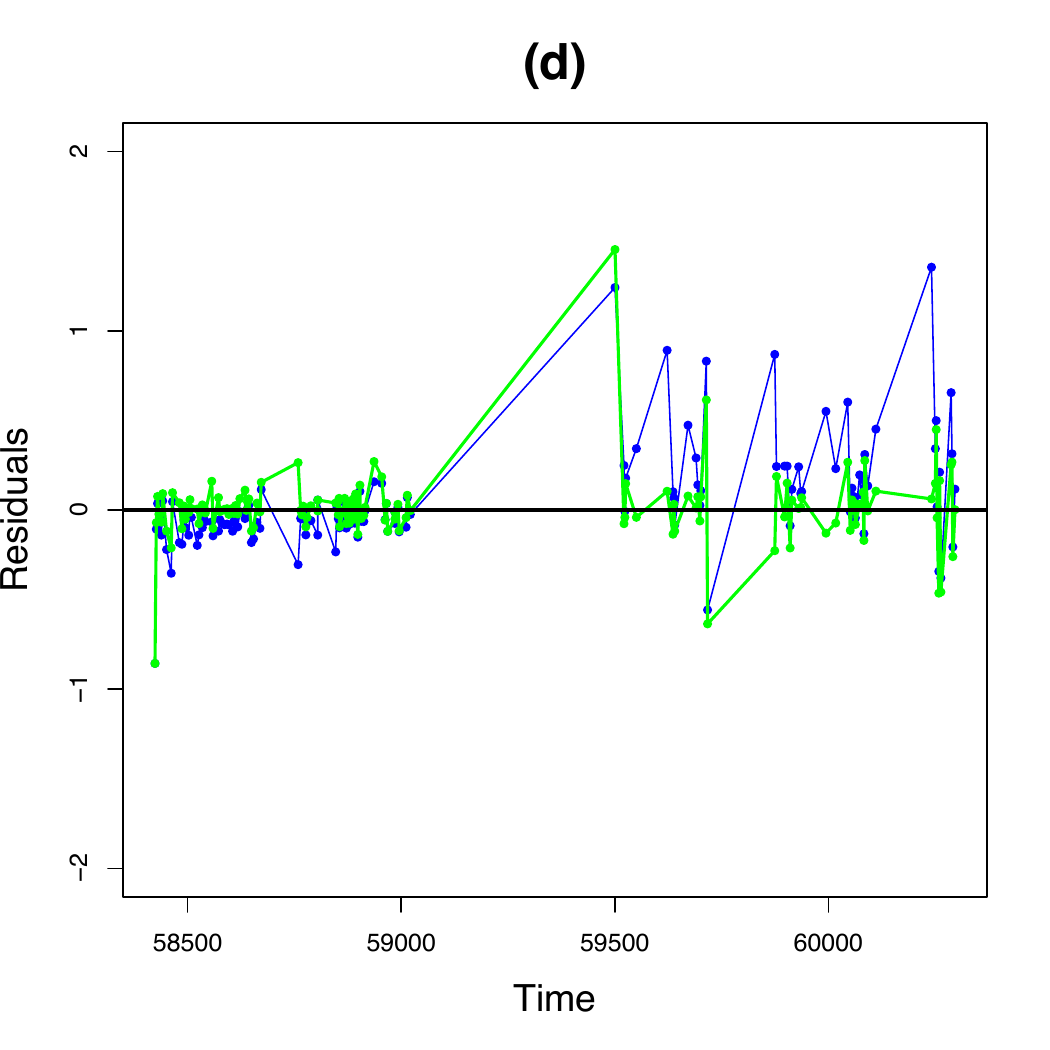}
\end{minipage}
\begin{minipage}{0.246\linewidth}
\includegraphics[width=\textwidth]{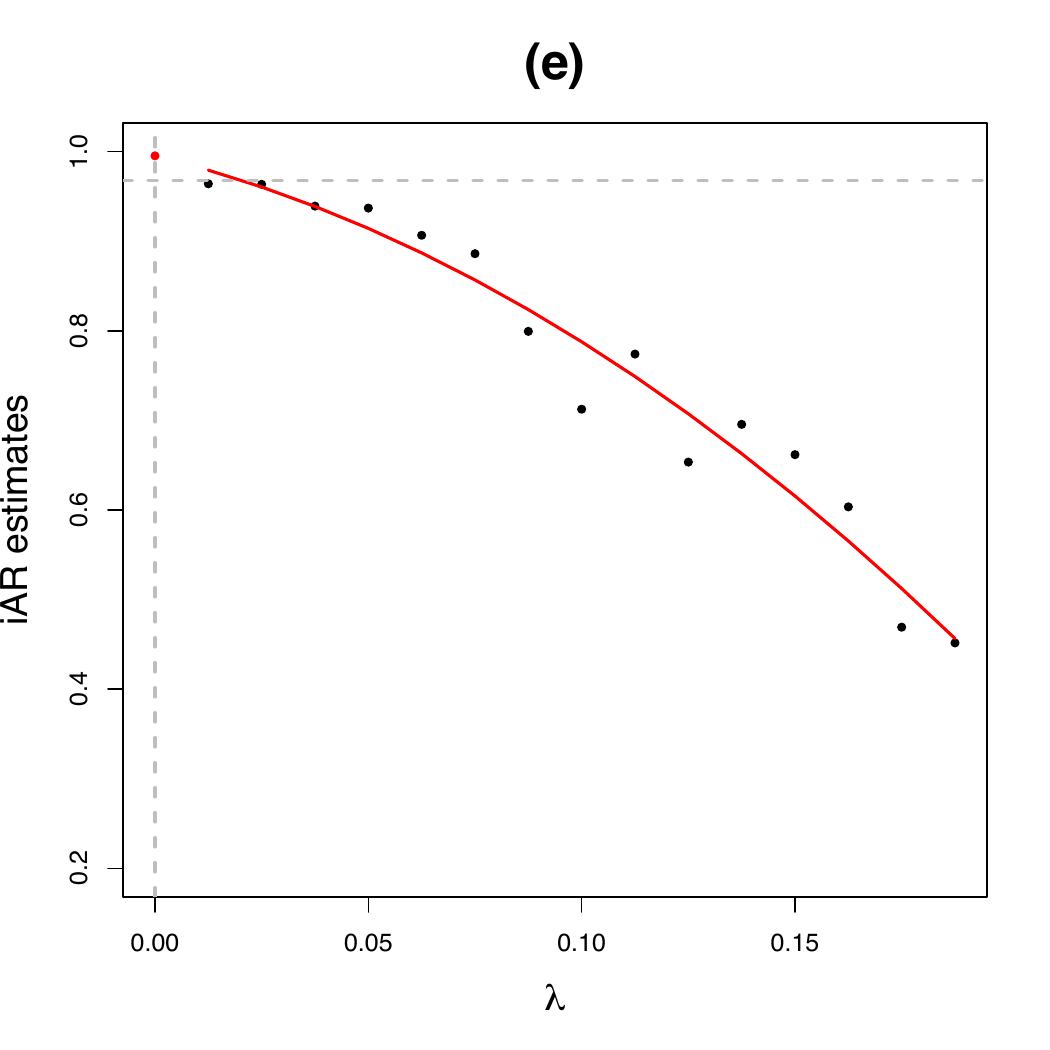}
\end{minipage}
\begin{minipage}{0.246\linewidth}
\includegraphics[width=\textwidth]{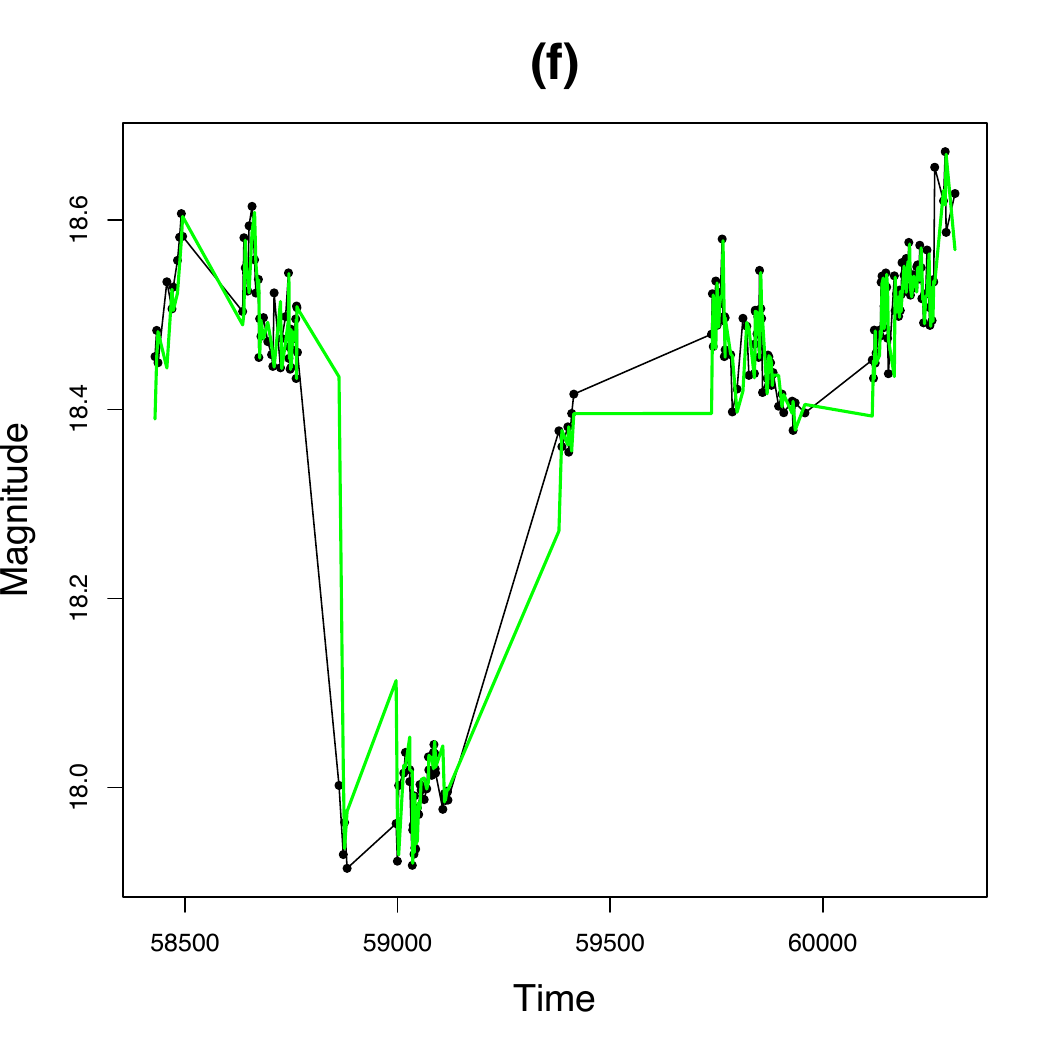}
\end{minipage}
\begin{minipage}{0.246\linewidth}
\includegraphics[width=\textwidth]{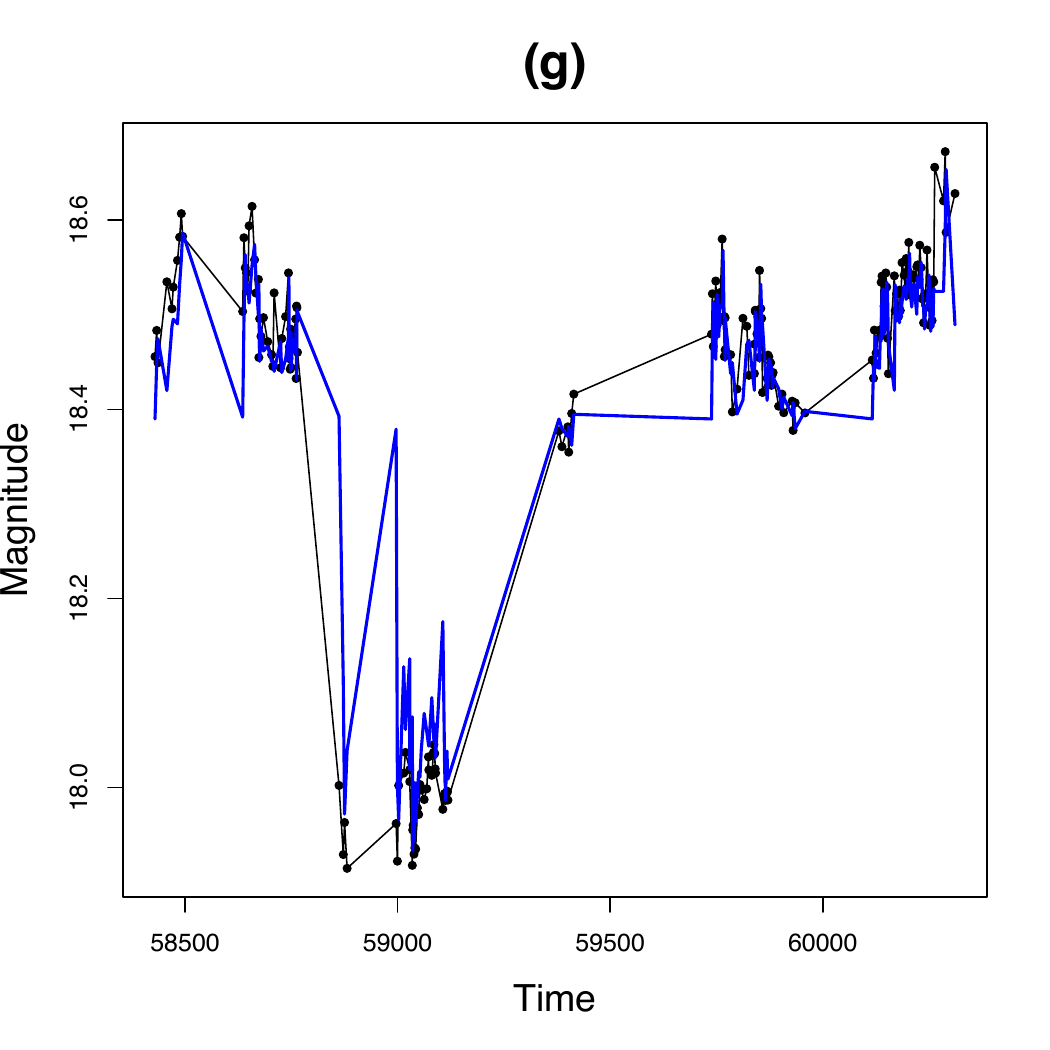}
\end{minipage}
\begin{minipage}{0.246\linewidth}
\includegraphics[width=\textwidth]{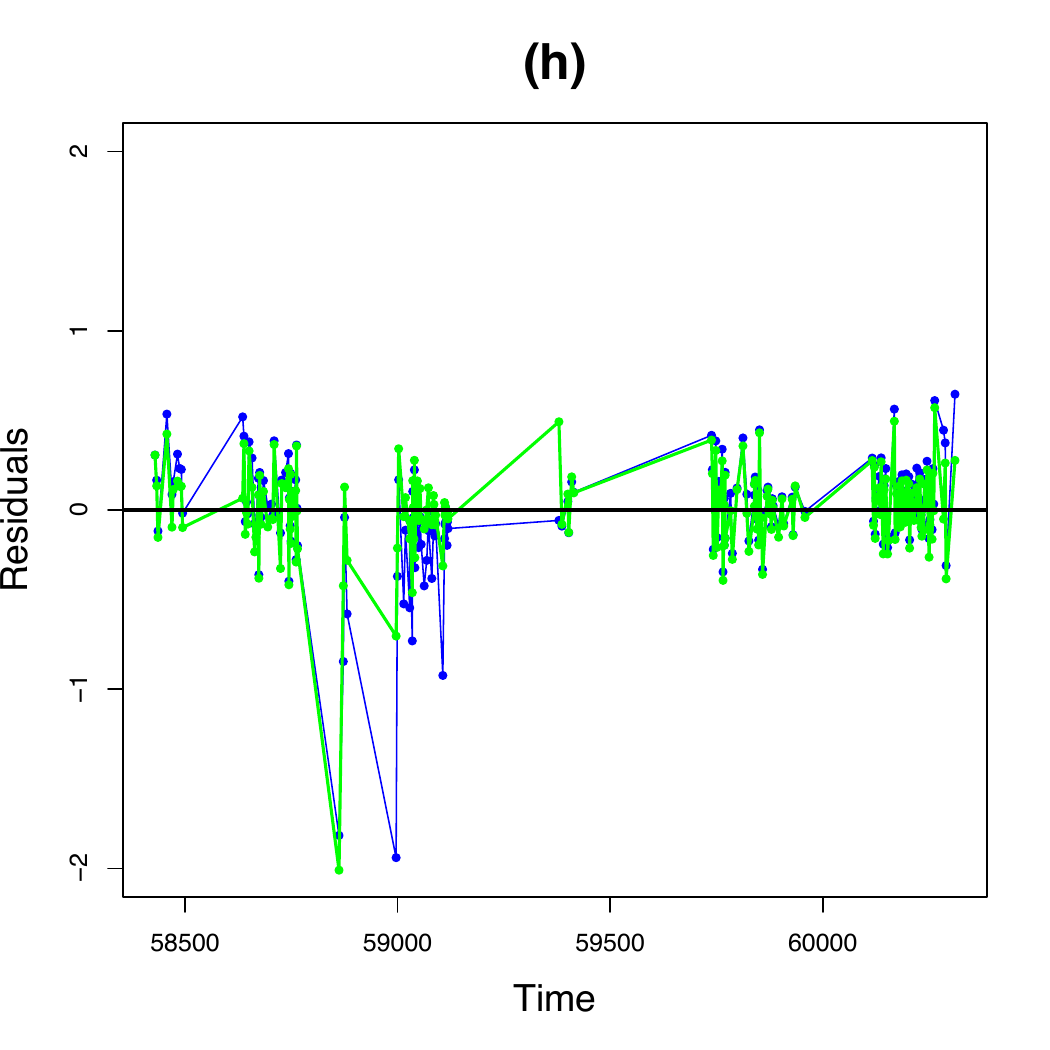}
\end{minipage}
\caption{Two examples of AGN light curves observed by ZTF survey fitted by the iAR model using both SIMEX and MLE estimation procedures. Figures a) and e) illustrate the SIMEX estimation of the $\phi$ parameter of the iAR model in both examples. Figures b) and f) show the fit of the iAR model using the SIMEX estimate of the autocorrelation parameter and Figures c) and g) show the fit of the iAR model using the MLE estimate of this parameter. Finally, Figures d) and h) show the residuals of the iAR model fits for both estimation methods. In these figures, black line is the observed light curve. green and blue lines are the fitted (residuals) values obtained by the SIMEX and MLE estimation procedures respectively.\label{fig:ZTFexamples}} 
\end{figure*}

When focusing on particular examples, we can observe light curves where the SIMEX method significantly improves the fit to the data. For instance, Figures \ref{fig:ZTFexamples} a)-d) show the fit of the iAR model using both the SIMEX and MLE estimation methods for the AGN light curve from alert object ``ZTF18acbzncb'' observed in the g-band by ZTF survey. For this light curve, the $\tau$ parameter estimated by MLE is $82.83$, while using the SIMEX method with parameters according to the Setting 2 mentioned previously yields a $\tau$ estimate of $368.24$. The difference obtained in the estimation of this parameter implies a reduction in the MSE obtained from the SIMEX estimation (0.033)  compared to the MLE estimation (0.056). The improvement in the fit of the SIMEX estimation is most evident in the second half of the light curve, where Figure c) shows that the fit of the MLE estimation is below the true values. This is corroborated in Figure d) where the residuals obtained with both estimation methods are compared.\\ 

A second example is presented in Figures \ref{fig:ZTFexamples} e)-h). This QSO light curve was also observed in the g-band by ZTF survey from alert object ``ZTF18aceivsg''. As in the previous example, the SIMEX estimation provides a higher estimate for the $\tau$ parameter compared to the MLE estimation. Specifically, the $\tau$ estimate obtained with the SIMEX method was $213.18$, while the MLE estimate was $30.50$. Additionally, the light curve fit obtained using the SIMEX estimation outperformed that from the MLE estimation, with MSE values of $0.059$ and $0.094$, respectively.\\


\section{Conclusions}

In this article, we introduced a novel simulation-extrapolation (SIMEX) algorithm for estimating the autocorrelation parameter in autoregressive models of order one. We considered regular and irregular autoregressive models with and without additive noise. Through extensive simulations, we demonstrated that standard estimation methods, such as maximum likelihood (MLE) and least squared error (LSE), can yield biased estimators of the autocorrelation parameter when the true value is close to one. This is a statistical problem known as the unit root problem, which refers to the fact that the parameter is close to the boundary of the parameter space. In this case, the boundary corresponds to a value of one.  When $\phi=1$, the process is non-stationary and unstable. As a consequence, it is not possible to obtain a reliable estimator. When $\phi$ is close to $1$, computer algorithms have difficulty distinguishing the true value of the parameter due to accuracy constraints. This root proximity problem complicates estimation, as the true value may be extremely close to the upper bound of the parameter space $(0, 1)$ and, in many cases, the algorithm fails to produce estimates, making it difficult to assess the variance of the estimator.\\

Monte Carlo experiments performed in this work show that the SIMEX estimation methodology improves substantially the estimation of the autoregressive model under both regular and irregular sampling, especially under challenging conditions such as near-unit root scenarios or the presence of additive noise. Monte Carlo simulations revealed substantial reductions in bias and improvements in estimation precision compared to standard methods like MLE and LSE. However, above certain thresholds, all method begin to fail by large factors.\\

We further show that the novel SIMEX algorithm offers an alternative for reliable estimation of the damping timescale parameter in both simulated and real AGN light curves observed in the ZTF survey. In both empirical applications, the proposed SIMEX method with time dependence methodology works well, validating the practical utility of this approach, yielding better model fits and lower MSE compared to the other estimation techniques. These results highlight the SIMEX method ability to provide more reliable estimates of the damping timescale parameter ($\tau$), which is essential for understanding AGN physical properties.\\

The SIMEX method proposed in this work aims to improve the variability estimation of AGN compared to the maximum likelihood estimator. In the context of autoregressive models for irregular time series, there are few works that use the maximum likelihood estimation (for example, the iAR model). Generally, the variability estimation of AGN is performed using DRW models based on Markov Chain Monte Carlo (MCMC) methods with nested sampling. It is important to note that the iAR and DRW models are equivalent (under Gaussianity). Therefore, a less biased estimator of the $\phi$  parameter of the iAR model also allows to have a better estimator of the $\tau$ parameter and consequently to improve the quality of fit of the DRW model.  The decision to implement the SIMEX methodology on the iAR model (and not directly on the DRW model), is due to the fact that this model can be fitted with less computational cost. The performance of the MCMC regarding bias is of out the scope of this work. We implemented this methodology in non-Bayesian methods.\\

As mentioned in Section \ref{sec:intro}, the estimation of the relaxation time from these models (iAR or DRW) assumes that this parameter is constant over time. Exploring this limitation is outside the scope of this work. However, it can be approached from a sequential estimation of the parameter $\phi$ of the iAR model (see for example, \cite{Elorrieta_2023}). In a future work, an adaptation of the SIMEX estimator for time-varying autoregressive parameter will be considered. \\



\section*{Acknowledgments}

The authors acknowledge support from the ANID – Millennium Science Initiative, AIM23-0001 awarded to the Millennium Institute of Astrophysics MAS (FE, WP, SE, FEB). FEB acknowledges support from ANID-Chile BASAL CATA FB210003, FONDECYT Regular 1241005.

\bibliographystyle{aa}
\bibliography{example}

\end{document}